 \documentclass[final,3p,times,authoryear]{elsarticle}

\usepackage{amssymb, amsmath, color,graphicx,bm,soul}
\usepackage[percent]{overpic}
\usepackage[section]{placeins}

\newcommand{\bfC}{\boldsymbol{C}}%
\newcommand{\bfsigma}{\boldsymbol{\sigma}}
\newcommand{\bfepsilon}{\boldsymbol{\epsilon}}
\newcommand{\bfxi}{\boldsymbol{\xi}}

\journal{Journal of the Mechanics and Physics of Solids}

\begin{document}

\begin{frontmatter}

\title{Morphometric description of strength and degradation in porous media}

\author[Duke]{A. Gu\'evel}
\corref{mycorrespondingauthor}
\cortext[mycorrespondingauthor]{Corresponding author}
\ead{alexandre.guevel@duke.edu}

\author[Louvain]{H. Rattez}

\author[Duke]{E. Veveakis}

\address[Duke]{Civil and Environmental Engineering Department, Duke University, Durham, NC 27708-0287, USA}

\address[Louvain]{Institute of Mechanics, Materials and Civil Engineering (IMMC), Universit\'e Catholique de Louvain, B-1348 Louvain-la-Neuve, Belgium}

\begin{abstract}
The influence of the microstructural geometry on the behavior of porous media is widely recognized, particularly in geomaterials, but also in biomaterials and engineered materials. Recent advances in imaging techniques, such as X-ray microcomputed tomography, and in modeling make it possible to capture the exact morphometry of the microstructure with high precision. However, most existing continuum theories only partially account for the morphometry. We propose here a unifying approach to link the strength of porous materials with the necessary and sufficient microstructural information, using Minkowski functionals, as per Hadwiger's theorem. A morphometric strength law is inferred from synthetic microstructures with a wide range of porosities and heterogeneities, through qualitative 2D phase-field simulations. Namely, the damage is modeled at the microstructural level by tracking the solid-pore interfaces under mechanical loading. The strength is found to be best described by an exponential function of the morphometers, thus generalizing early works on metals and ceramics. We then show that the predictiveness of this relationship extends to real porous media, including rocks and bones.
\end{abstract}

\begin{keyword}
porous media, morphometry, strength, phase-field modeling, degradation, damage, Minkowski functionals
\end{keyword}


\end{frontmatter}


\section{Introduction}

Porous media represent a wide range of materials but also a tremendous challenge to be fully understood and harnessed. Among them, geomaterials, stemming from millions of years of transformations under harsh conditions, represent a particularly complex subclass, inasmuch as processes in these media are multiphysics and multiscales. Recent advances in geosciences found, however, that this complexity may boil down to the great heterogeneity and stochasticity of geomaterials' microstructures. For instance, pressure solution (\cite{Niemeijer2009,Croize2013,VandenEnde2019,Guevel2020}), strain localization (\cite{Vardoulakis1995, Kawamoto2018}), frictional instabilities (\cite{Rattez2018a, Rattez2018b}), fault reactivation (\cite{Veveakis2014,Lesueur2020}), and granular flow (\cite{Buscarnera2021}) largely depend on the microstructural geometry, or \textit{morphometry}. The same conclusion holds for engineered porous materials as well, such as ceramics (\cite{Salvini2018}) and energetic materials (\cite{Chun2020}). Biomaterials, forming a third subclass of porous media (\cite{Huyghe2002}), are also increasingly studied in the light of their microstructures, in particular in bones mechanics (\cite{Augat2006,Wachter2002}).

Therefore, modeling the microstructural dynamics is a crucial step towards better understanding the macroscopic behavior of porous media. In metallurgy, this has been achieved with phase-field modeling (\cite{Allen1979, Provatas2010, Bhattacharyya2019}), which tracks the grains' interfaces in finite elements, and in geosciences, with Discrete Element Modeling (\cite{OSullivan2012,Kawamoto2018}), which tracks the grains as discrete elements, and recently also with phase-field modeling (\cite{Guevel2020}). That said, due to computational limitations, relying solely on microscopic simulations for large scales is unrealistic. On the other hand, the existing constitutive macroscopic laws rely heavily on destructive experimental calibration, which can be unfeasible because of the limited availability of materials or the impossibility to reproduce the environmental conditions. For example, obtaining rocks from high depths can be prohibitively expensive or even impossible, and geological time scales are hardly reproducible in laboratory. As for biomaterials, such as bones, they are best studied in vivo (see \cite{Fragogeorgi2019} e.g.).

The correlation between the strength of materials and their microstructural geometry was first theorized, through the foundation of poromechanics, by \cite{Biot1941}, and for soils specifically, by \cite{Terzaghi1943}. This effort was intensified following the catastrophic failure of welded Allies ships through brittle fracture during the Second World War, with focus on metals and ceramics. The strength of brittle polycrystalline metals was quantitatively addressed by \cite{Orowan1949}, proposing what would ultimately become the Hall--Petch effect (\cite{Hall1951, Petch1953}). The Hall--Petch relationship asserts that the yield stress, or similarly the flow stress, scales as the inverse square root of the mean grain size. However, \cite{Li2016a} recently showed, upon critically reviewing a large amount of experimental data, that other types of laws, including inverse exponential, turn out as satisfying. Furthermore, \cite{Ryshkewitch1953} experimentally found that the strength of porous brittle polycrystalline materials, such as ceramics, is an inverse exponential function of the porosity. This was confirmed more recently with porosity-controlled experiments and modeling by \cite{Liu2017}. Building upon the previous results from non-porous and porous brittle materials, \cite{Knudsen1959} inferred a multiplicative dependence of the strength on grain size and porosity (Eq.~(7) therein), which was checked experimentally for ceramics at different temperatures. These efforts culminated to the advent of modern poromechanics (see \cite{Coussy2004,Dormieux2006}), reconciling continuum mechanics with Biot's theory, and relevant to a wide range of disciplines, including geo, bio and material sciences. However, the strength of porous media was mostly studied with respect to porosity and grain size. According to morphometry theory and in particular Hadwiger's theorem (see \cite{Armstrong2019} and Section \ref{subsec: morphometry}), two more descriptors of the microstructure, which we refer to as \textit{morphometers}, are needed to fully account for the morphometric effects. On these bases, we propose to investigate a morphometric strength law of the form:
\begin{equation}
	\label{eq:strength_law}
	\sigma_f = \sigma_f^* f(M_0,M_1,M_2,M_3)
\end{equation}
where we choose to measure the strength via a flow stress $\sigma_f$ (see Section \ref{subsec:measure_stength}), the $M_i$ (i=0,\dots,3) denote the four required morphometers, the star quantities are reference values depending on the material and environmental conditions, and $f$ is a multiplicative function of the $M_i$, following \cite{Knudsen1959}. In particular, $M_0$ denotes the porosity and $M_2$ the grain size. Our objective is to investigate the possible application of this law to porous materials in general, including geomaterials and biomaterials. Thereby, we subsume the microscopic mechanisms responsible for failure under the generic umbrella of damage. For geomaterials, this encompasses debonding, dissolution, cracks and breakage. Recent investigations in geosciences on the influence of the morphometry notably include the work of \cite{Zhang2016}, \cite{Wetzel2021} and \cite{Buscarnera2021}. For biomaterials, in particular for bones, damage comprises dissolution, as in osteoporosis, and cracks, particularly in cortical bones (\cite{Augat2006}). While the focus in biomechanics is mostly on the relationship between elastic moduli and morphometers, some results corroborate the conclusions inferred for the previous subclasses of porous materials, such as the exponential relationship between strength and porosity (see \cite{Wachter2002} and \ref{app:strength bones}). However, a mathematically-consistent law such as Eq.~(\ref{eq:strength_law}) has not been proposed yet, accounting for the necessary and sufficient microstructural information.

In order to numerically investigate the dependency of the strength of porous materials on the morphometers and specify Eq.~(\ref{eq:strength_law}), we control the variability of the morphometry by using synthetic microstructures (SMs). This also allows to run a sufficiently large ensemble of simulations to be statistically meaningful and account for the stochastic nature of real microstructures (\cite{Chun2020}). Interest in synthesizing microstructures is intensifying in material sciences, owing to recent computational advances. In particular, machine learning enables to create highly realistic, tunable microstructures (\cite{Mosser2017,Chun2020}). While the latter represents the state of the art, we have used, as a first step, a more straightforward and much less computationally-demanding approach (see Section \ref{subsec:synthetic_microstructures}).

In all, we aim here at bridging the gap between microscopic and macroscopic modeling, by calibrating macroscopic laws with morphometric parameters upscaled from simulations on synthetic microstructures, hinging, as much as possible, on non-destructive methods. We should then check that the predictiveness of this relationship extends to real porous media,  by using CT scans. While complex multiphysics processes like pressure solution may still require explicit modeling of the microstructure (\cite{Guevel2020}), we suggest that as far as the influence of the microstructural geometry is concerned, it is possible to upscale only the essential morphometric information.

\section{Methods}
\label{sec: Methods}

\subsection{Upscaling of the microstructural information}
\label{subsec: morphometry}

The morphometers $M_i$ in Eq.~(\ref{eq:strength_law}) are appropriately described by the Minkowski functionals of the domain formed by the grains $\Omega$ (\cite{Armstrong2019}). Hadwiger's theorem guarantees that a microstructure is fully described by $d+1$ Minkowski functionals, where $d$ is the microstructure's dimension, in the sense that any other descriptor that is additive, motion-invariant and conditionally continuous would be a linear combination of those functionals (see \cite{Hadwiger1957} and also \cite{Klain1995} for a short proof). In 3D, the 4 needed Minkowski functionals are (\cite{Armstrong2019})
\begin{equation}
	M_0(\Omega) = \int_\Omega \text{d}V,
\end{equation}
the total volume of the grains, 
\begin{equation}
	M_1 (\Omega)= \int_{\partial\Omega} \text{d}S,
\end{equation}
their total surface area, 
\begin{equation}
	M_2 (\Omega)= \int_{\partial\Omega} \left( 1/r_1+1/r_2 \right)\text{d}S,
\end{equation}
their total mean curvature, where $r_1$ and $r_2$ denote the principal radii of curvature of the surface element $\text{d}S$, and
\begin{equation}
	M_3(\Omega) = \int_{\partial\Omega} \left( 1/r_1r_2 \right)\text{d}S = 4\pi \chi(\Omega),
\end{equation}
their total Gaussian curvature, directly related to the Euler characteristic $\chi$ by the Gauss-Bonnet theorem (\cite{Armstrong2019}). In practice, we will use the porosity $n$ as a measure of $M_0$  and the mean grain size for $M_2$ , the latter approximating the inverse of the average mean curvature. In this contribution, we will restrict our results to $d=2$ as a first step, and therefore use only 3 morphometers to constrain the strength law Eq.~(\ref{eq:strength_law}).

\subsection{Synthetic microstructures}
\label{subsec:synthetic_microstructures}

\begin{figure}[h!]
	\centering
	\begin{overpic}[width=0.3\linewidth]{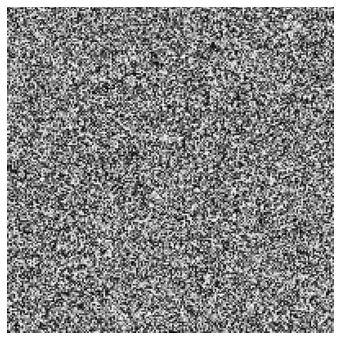}
		\put (-8,87) {\color{black} $a)$}
	\end{overpic}
	\hspace{0.1\linewidth}
	\begin{overpic}[width=0.3\linewidth]{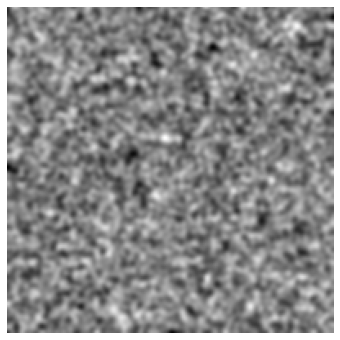}
		\put (-8,87) {\color{black} $b)$}
	\end{overpic}
	
	\vspace{0.01\linewidth}
	
	\begin{overpic}[width=0.3\linewidth]{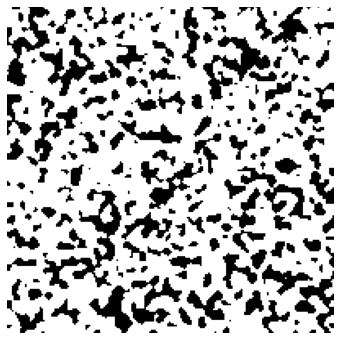}
		\put (-8,87) {\color{black} $c)$}
	\end{overpic}
	\hspace{0.1\linewidth}
	\begin{overpic}[width=0.3\linewidth]{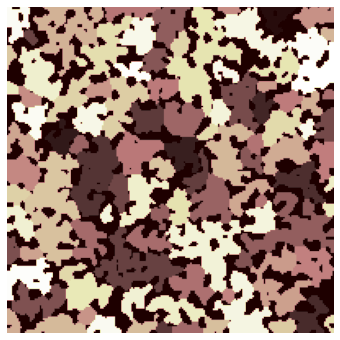}
		\put (-8,87) {\color{black} $d)$}
	\end{overpic}
	\caption{a) Random noise $R[x,y]$, i.e.\ array of random values between 0 and 1. b) Gaussian filter $F[x,y]$ applied onto the random noise, with a standard deviation of 1.76, or equivalently, $h=3$. c) Resulting binarized SM $B[x,y]$ upon applying the threshold $n=0.3$. d) SM segmented into 110 grains via watershed algorithm.}
	\label{fig:MS_generation}
\end{figure}

To find the function $f(M_i)$ in Eq.~(\ref{eq:strength_law}), we run a large amount of simulations on SMs, which are generated using the Python open-source package \texttt{PoreSpy} \footnote{http://porespy.org} (\cite{Gostick2019}). The starting point is a random noise (see Fig.~\ref{fig:MS_generation}a), that is, a 2D array $R[i,j]$ ($i=0,\dots,N-1$,$j=0,\dots,N-1$) of dimension $N$ of random values between 0 and 1. It is then convoluted with a Gaussian filter (see Fig.~\ref{fig:MS_generation}b)
\begin{equation}
    G[i,j]=e^{-\frac{i^2+j^2}{2s^2}},
\end{equation}
which standard deviation $s=\frac{N}{40h}$, or blurriness, is inversely proportional to a heterogeneity parameter $h$, and where 40 is a scaling factor; the heterogeneity of the output is then independent from the image size. The filtered output $F[i,j]$ is thus calculated as follows (see \cite{Gonzalez2008} e.g.):
\begin{equation}
	F[i,j]=R[i,j]*G[i,j]=\sum_{u=0}^{N-1} \sum_{v=0}^{N-1} R[u,v]G[i-u,j-v],
\end{equation}
where $*$ denotes the convolution operator. Finally, after uniformizing the blurred noise $F[i,j]$ between 0 and 1, the array is binarized into $B[i,j]$ (see Fig.~\ref{fig:MS_generation}c) by using the porosity $n$ as the threshold, yielding the final SM. The two input parameters controling the generation of the SMs are thus the heterogeneity $h$ and the porosity $n$. We will see that $h$ controls the narrowness and skewness of the morphometers distributions. 
To restrict the study to realistic microstructures, $n$ is varied from $0.15$ to $0.4$ and $h$ from $2$ to $5$, creating an ensemble of 42 SMs (see \ref{app:summary_table}). This ensemble forms the training batch, from which the dependence of the flow stress on the morphometry will be inferred. Throughout this work, the dimensions of the synthetic SMs are restricted to $200\times200$ pixels, as a compromise between sufficiently realistic microstructures and reasonable mesh sizes, proportional to this resolution. It is then possible to obtain the distributions of the morphometers $M_0$, $M_1$, $M_2$, i.e.\ of the pore size, perimeter and grain size, respectively, starting with the marker-based watershed segmentation algorithm introduced by \cite{Gostick2017}. The values of the filter size and of the Gaussian blur therein must be chosen carefully to maximize the number of segmented regions (grains or pores). In particular, since the optimal values for those parameters vary with the grain/pore size, they are expected to vary with $h$. Namely, we found that the filter size varies from 2 to 9 pixels for our range of SMs, whereas the blur varies from 0.2 to 0.8. Upon segmenting the pores, their sizes are obtained by determining the equivalent disks for every pores. Upon segmenting the grains, their perimeter is obtained via an edge-finding algorithm optimized for fast computing introduced by \cite{Benkrid2000}. The grain sizes are obtained similarly to the pore sizes. The last morphometer $M_3$, the Euler characteristic, is not concerned by distributions since it is a topological quantity. While the distributions of $M_0$ and $M_1$ are useful when comparing the SMs with real microstructures (see Section \ref{subsec:resemblence_with_real_microstructures}), only the grain size distribution is calculated for every SMs, in order to determine the mean grain size $M_2$. The latter is taken as the mean of the lognormal fit of the grain size distribution, as explained in Section \ref{subsec:resemblence_with_real_microstructures}. $M_0$ and $M_1$ are directly obtained from calculating the porosity, i.e.\ the ratio of the black pixels over the total number of pixels, and the total perimeter, respectively. Finally, the Euler characteristic $M_3$ is calculated from an integral geometry formula in discretized space. In the current 2D restriction, $M_3$ is simply the number of objects minus the number of holes. We refer to the documentation of \texttt{PoreSpy} and \texttt{skimage} available online for further details (see \ref{app:calculation_morphometers}).

\subsection{Damage phase-field modeling of the microstructure}
\label{subsec: phase-field modeling}

The evolution of the microstructure is modeled via a phase field $\varphi$ differentiating the pores from the grains, as introduced in \cite{Guevel2020}, and describing the damage of the grains. As such, the phase field measures the changes of morphometry, and thus controls the evolution of the morphometers (see Section \ref{subsec:numerical_output}). For all simulations, we use the open-source multiphysics finite-element platform \texttt{MOOSE} (\cite{Permann2020}). Following \cite{Guevel2020} and assuming that the grains cannot heal, we consider the following Allen--Cahn equation, coupled with elastic mechanics:
\begin{equation}
	\label{eq:system_equations}
	\begin{cases}
		\tau \dot\varphi = \kappa\triangle\varphi - G g'(\varphi) - \lambda(\bfepsilon)h'(\varphi), \\ 
		\nabla \cdot \bfC(\varphi)\bfepsilon =0,
	\end{cases}
\end{equation}
where $\tau$ is the relaxation time, $\kappa$ the interfacial energy coefficient, $G$ the double-well's height, $\bfepsilon$ the strain tensor, $\triangle$ and $\nabla \cdot$ denote the Laplacian and divergence operators, the dot denotes the time derivative and the prime the derivative with respect to $\varphi$. The polynomials $g(\varphi)=\varphi^2(1-\varphi)^2$ and $h(\varphi)=\varphi^2(3-2\varphi)$ correspond to a double-well potential and an interpolation function, respectively. Furthermore, $\lambda(\bfepsilon)=1/2 (\bfC_g-\bfC_p)\bfepsilon \cdot \bfepsilon$ denotes the difference of elastic energy between the grains and pores phases. Therein, $\bfC_g$ and $\bfC_p$ denote the elastic tensor of the grains and the pores, respectively. Finally, $\bfC(\varphi)=(1-h(\varphi)) \bfC_p + h(\varphi) \bfC_g$ is the elastic tensor of the mixture. In doing so, we apply the Voigt-Taylor homogenization scheme (see \cite{Ammar2009} e.g.), which consists in interpolating the partial stresses of the grains and pores phases while assuming homogeneous strains, thus recovering a familiar poromechanics formulation. The damage of the grains occurs when the elastic energy is large enough to tilt the double-well potential beyond the saddle-node bifurcation point (\cite{Guevel2020}). 

This model stems from the coupling of the microscale mechanics (at the grains scale) and the macroscale mechanics (mixture of grains and pores). Namely, Eq.~(\ref{eq:system_equations})-1 stems from the micro-momentum balance (\cite{Fried1993, Gurtin1996})
\begin{equation}
	\nabla\cdot\bfxi+\pi=0,
\end{equation}
where $\pi$ is the microforce (scalar), energy-conjugate of $\varphi$, $\bfxi$ is the microstress (vector), energy-conjugate of $\nabla\varphi$, whereas Eq.~(\ref{eq:system_equations})-2 corresponds to the usual macro-momentum balance. They are coupled via the \textit{generalized relaxation equation} (\cite{Guevel2020}), equivalent to the second law of thermodynamics as described by the following dissipation inequality
\begin{equation}
	-\dot\psi - \pi\dot\varphi + \bfxi \cdot \nabla\dot\varphi - \bfsigma:\dot\bfepsilon \leq 0,
\end{equation}
where $\psi$ denotes the free energy, described in its usual Landau form (\cite{Landau1937})
\begin{equation}
    \psi(\varphi,\nabla\varphi,\bfepsilon) = Gg(\varphi) + (1-h(\phi))\frac{1}{2}\bfepsilon \cdot \bfC_p \bfepsilon + h(\varphi)\frac{1}{2}\bfepsilon \cdot \bfC_g \bfepsilon + \frac{\kappa}{2}\lvert\nabla\phi\rvert^2,
\end{equation}
and $\bfsigma=\bfC(\varphi)\bfepsilon$ is the Cauchy stress tensor, resulting from the interpolation of the partial stresses. Therein, the state variables are $\varphi$, describing the normal variations of the grains interface, $\nabla\varphi$, describing the tangential variations of the grains interface, and the macroscopic (elastic) strain tensor $\bfepsilon$. For simplicity, we consider $\nabla\varphi$ and $\bfepsilon$ to be non-dissipative, thus neglecting viscous effects; the viscous effect associated with the phase field would delay phase changes as shown in \cite{Guevel2020}. 

Here, damage is not specified but includes dissolution, debonding and microcracks. In particular, the final system of equations (\ref{eq:system_equations}) is similar to the one used for modeling fractures at the continuum scale (see \cite{Kuhn2010} e.g.). The fundamental difference is that our model is applied directly at the grains scale, whereas phase-field damage models have so far been applied to the continuum scale. Hence, instead of the usual differentiation between the intact and the damaged phases, our model differentiates between the grains and the pores phases, where the growth of the pores phase at the expense of the grains phase embodies the microstructural damage. We distinguish the latter from its upscaled manifestation at the continuum scale, which we call the \textit{degradation} (see Section \ref{subsec:numerical_output}). In particular, while the damaged phase in the macroscopic theory is not energetic, here, the pores phase is associated with an elastic energy (albeit not allowing shearing), yielding a mixture of elastic energies, as described above. Therefore, the only differences in the form of our phase-field equation with the continuum damage phase-field modeling are the presence of an elastic energy for the phase $\varphi=0$ and the degrees of the polynomials representing the potential $g(\varphi)$ and the interpolation function $h(\varphi)$. Our model can also be compared to the one used for grain growth in (non-porous) metals under mechanical loading (see \cite{Tonks2011} e.g.), where each grain of a polycrystal is represented by an order parameter. Unlike for metals, the pores phase in porous media such as geomaterials plays a crucial role, so that it is given an independent role in our model. The main difference between modeling porous media and metals is thus that we discriminate the pores from the skeleton, without differentiating, as a first step, the different grains, inasmuch as they are not necessarily distinguishable within the skeleton, depending on the materials; for instance, as discussed in the following, unlike a sandstone, a cortical bone can hardly be decomposed into grains. In practice, this translates into dropping the grain-to-grain interfacial energy term present in the polycrystals phase-field model. In all, modeling the damage of the microstructure of porous media is performed here as a combination of the concepts used in continuum damage and polycrystals phase-field modeling.

Upon choosing a characteristic length $l_0$, a characteristic time $t_0=\tau/G$ and a characteristic specific energy $G=1GPa$, we write (\ref{eq:system_equations}) in dimensionless form as follows:
\begin{equation}
	\label{eq:system_equations_dimensionless}
	\begin{cases}
		\dot\varphi = \hat\kappa \triangle\varphi - g'(\varphi) - \hat\lambda(\bfepsilon)h'(\varphi), \\ 
		\nabla \cdot \hat\bfC(\varphi)\bfepsilon =0,
	\end{cases}
\end{equation}
where $\hat\kappa = \kappa/G l_0^2$ is the (dimensionless) interfacial group, $\hat\lambda=\lambda/G$ is the activation energy group and $\hat\bfC=\bfC/G$. The dimensionless derivatives are noted similarly and the hat notation for dimensionless quantities is dropped in the following. The evolution of the damage via $\varphi$ is fully determined by $\lambda$, upon fixing $\kappa$ to an appropriately small value (see Section \ref{subsubsec: influence of kappa}). We also restrict our attention for simplicity to 2D problems, so that the elastic energy $\lambda$ is fully determined by the two Lam\'e parameters. As explained in \cite{Guevel2020}, the grains are considered as solids of very low porosity, and the pores as shear-free solids that are much more deformable. Namely, for geomaterials, we choose the first and second Lam\'e parameters of the grains both to be $30GPa$, and for the pores $1GPa$ and 0, respectively. Further major assumptions include small strains and that the assembly of grains considered is representative of the material. To focus on the effect of the microstructure on the strength, we will first perform simulations in tension with fixed lateral boundaries. Then, to determine the effect of the confinement, we will perform axisymmetric biaxial loadings. Unless mentioned otherwise, the stress used in output throughout this work is the vertical stress averaged on the top boundary of the domain. The associated strain is the vertical displacement of the top boundary divided by the initial length of the square domain.

\section{Model's performance}

\subsection{Measure of the strength}
\label{subsec:measure_stength}

\begin{figure}[h!]
	\centering
	\includegraphics[width=0.8\linewidth]{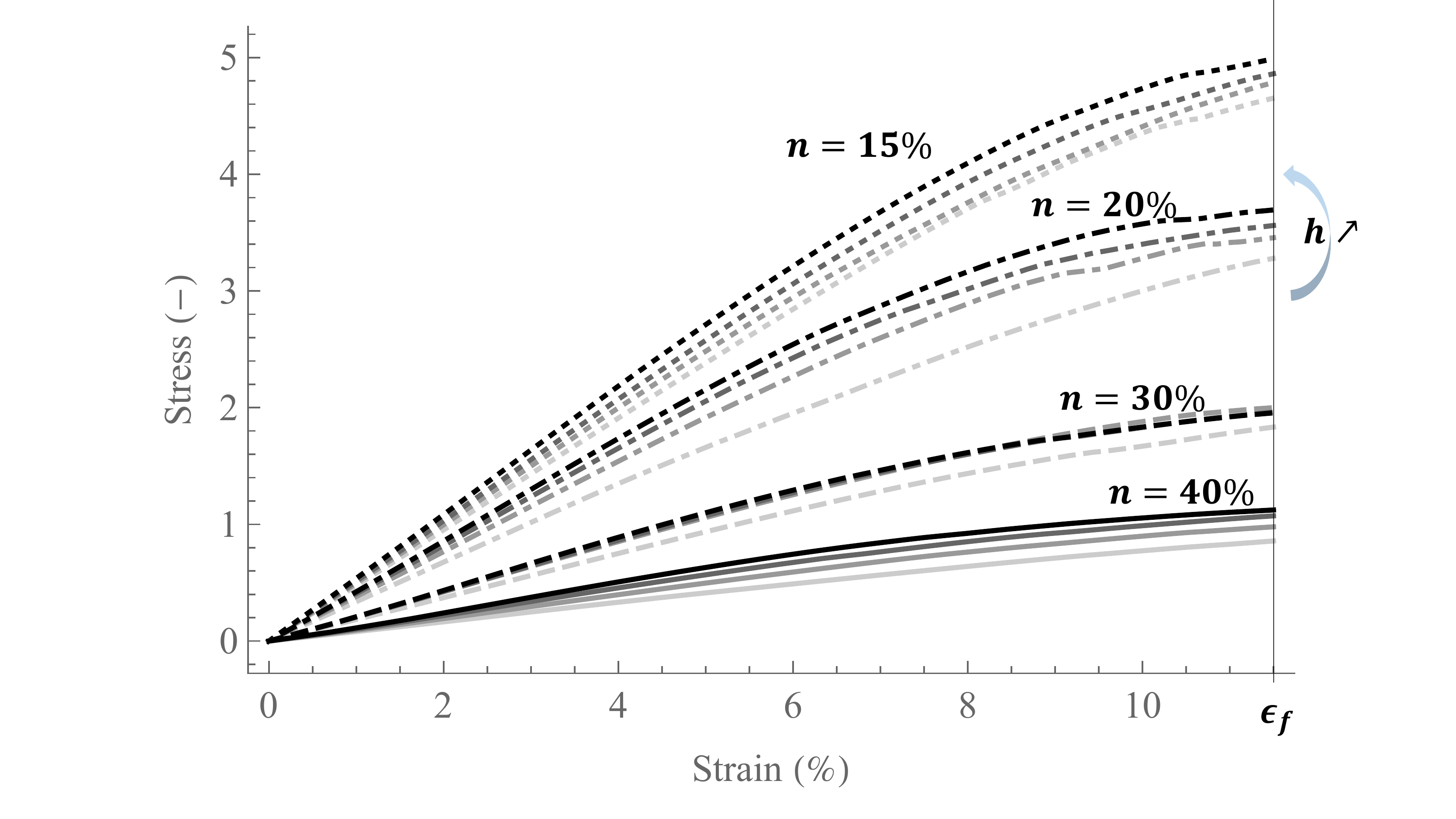}
	\caption{Stress-strain responses for 16 of the 42 SMs ($n \in \{0.15,0.20,0.30,0.40\}$, $h \in \{2,3,4,5\}$) until $\epsilon_f=11.5\%$, showing that all the SMs reach post-yield for this value. Changes from dotted to full lines correspond to an increase in porosity $n$, and increases in grey intensity to an increase in heterogeneity $h$. The rest of the SMs corroborates this assertion but are not shown here for visibility purposes.}
	\label{fig:determination_flow_stress}
\end{figure}

In solid mechanics, the word "strength" has a broad meaning. Most often, it corresponds to the maximum value that the stress can attain (peak stress). Alternative measures of strength include the yield stress, or even the value of stress at a given value of strain when the material is undergoing prolonged hardening (\cite{Sari2020}). Since the latter is the case in the present study, the strength is defined throughout this work as a value of the vertical stress for a certain value of a post-yield deformation attributed to a given material. The latter is determined for the SMs a posteriori, after obtaining the stress-strain output for all the training batch, as the minimum yield strain. Namely, as shown in Fig.~\ref{fig:determination_flow_stress} for 16 of the 42 SMs, we find that $\epsilon_f=11.5\%$ is the minimum strain required for all the SMs to enter yielding, that is, so that the stress-strain response is non-linear. The flow stress $\sigma_f$ is the stress corresponding to this value of $\epsilon_f$. It will be determined similarly for the real microstructures.

\subsection{Mesh convergence}

\begin{figure}[h!]
	\centering
	\includegraphics[width=0.6\linewidth]{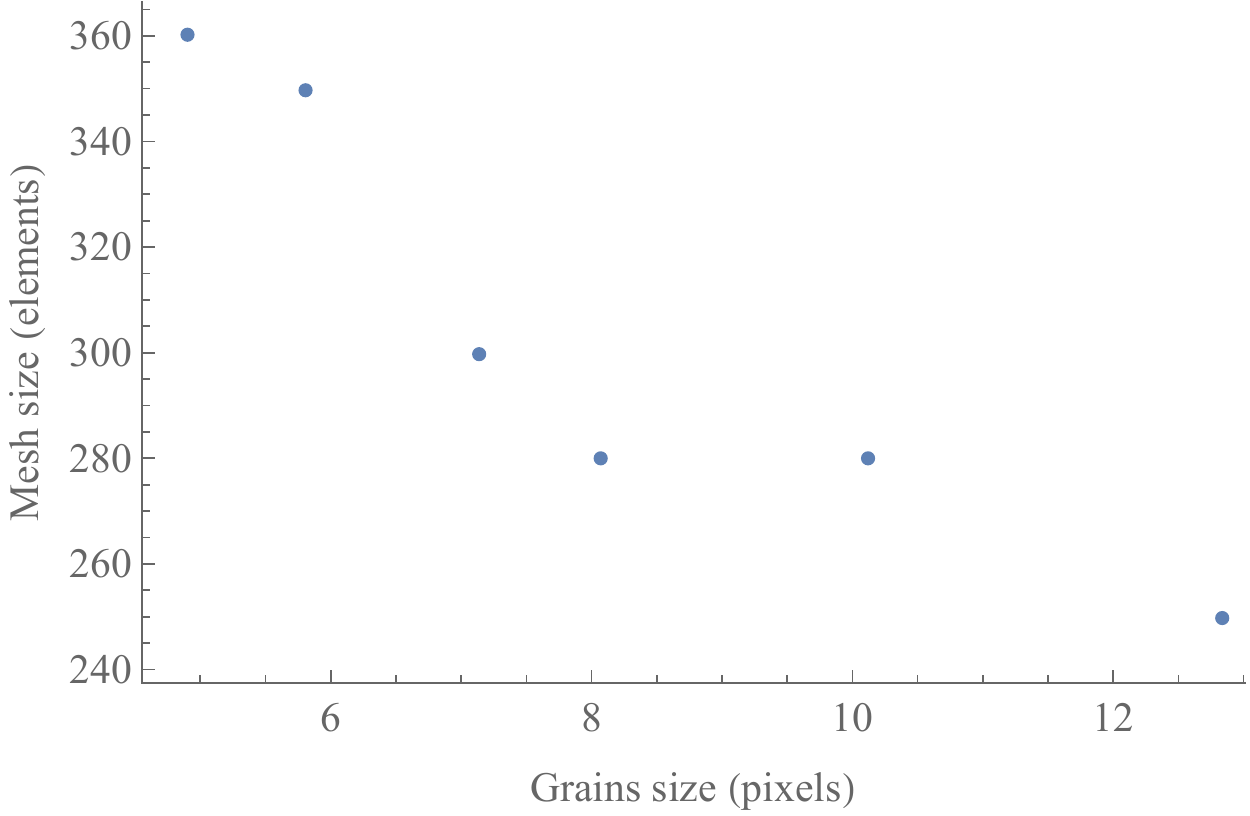}
	\caption{Evolution of the minimum mesh size (number of elements per side of the domain) required for mesh convergence with respect to the grain size.}
	\label{fig:mesh_conv}
\end{figure}

Before obtaining any meaningful results, it is important to study the mesh convergence. The initial conditions of the simulations are indeed determined by digitizing the input image, with respect to the chosen mesh dimensions. In particular, smaller microstructural dimensions require a finer mesh. Therefore, we expect to use finer meshes for SMs of smaller grain sizes. We refer to \ref{app:summary_table} for the summary of the grain sizes of our training batch. To assess the mesh convergence, we compare the values of the flow stress for the different SMs, to find that, indeed, the mesh dimensions increase with the grain size (see Fig.~\ref{fig:mesh_conv}). We then pick the minimum mesh size required for mesh convergence within a 2\% error. In practice, using Fig.~\ref{fig:mesh_conv}, we use the mesh sizes shown in Tab.~\ref{tab:mesh_size}. Note that the grain sizes are measured in pixels where $1px=l_0/N$, $l_0$ is the characteristic length of the problem taken as the image size and $N$ the image dimension. When $l_0$ is determined in the case of real microstructures, the length measurements will be given in $mm$ (or $\mu m$) in the following.
\begin{table}[]
	\centering
	\begin{tabular}{|c|c|}
		\hline
		Mean grain size (px) & Mesh size (elts) \\ \hline
		4-6                  & $360\times360$              \\ \hline
		6-7                  & $350\times360$            \\ \hline
		7-8                  & $300\times300$              \\ \hline
		8-13                 & $280\times280$              \\ \hline
	\end{tabular}
\caption{Number of elements used for the meshes in our simulations for different mean grain sizes.}
	\label{tab:mesh_size}
\end{table}

\subsection{Influence of the interfacial coefficient $\kappa$}
\label{subsubsec: influence of kappa}

\begin{figure}[h!]
	\centering
	\includegraphics[width=1\linewidth]{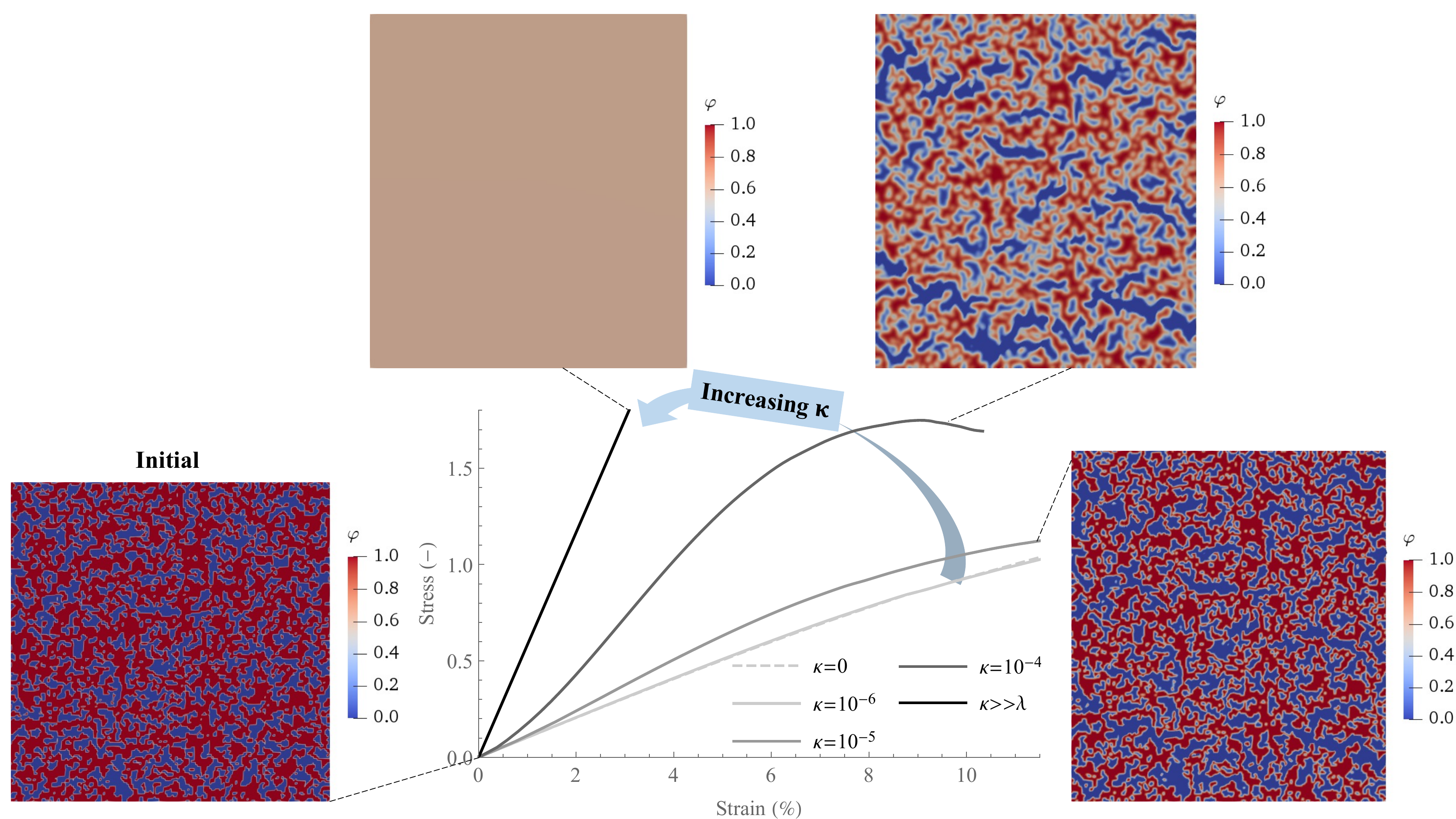}
	\caption{Mechanical response of the SM $(n=0.4,h=5)$ for different values of $\kappa$, including the limit cases $\kappa=0$ (sharp interface) and $\kappa \gg \lambda$ (pure diffusion). Note the spurious diffusion in the case $\kappa=10^{-4}$, leading to the closure of the smallest pores, as opposed to the case $\kappa=10^{-5}$.}
	\label{fig:effect_lap_coeff}
\end{figure}

While the activation energy $\lambda(\bfepsilon)$, coupled with the mechanical energy, is the main drive for the phase change, the interfacial coefficient $\kappa$ quantifies its diffusive character. When $\kappa \gg \lambda$, that is, when the activation energy is negligible compared to the interfacial energy, the process is purely diffusive, and therefore nonphysical in the current context. Conversely, when $\kappa \rightarrow 0$, the process falls back on the corresponding sharp interface problem, thereby losing the regularization provided by the diffuse interfaces, characteristic of phase-field modeling. We thus choose the value of $\kappa$ not too large to avoid spurious interfacial diffusion, but not too low to keep diffuse interfaces. The former would typically induce the closure of the smallest pores, independently from the mechanical loading. To do so, we perform simulations for different values of $\kappa$, while the microstructure and the other parameters are fixed. To be conservative, we choose the SM with the smallest mean grain size in our training batch, namely for $(n=0.4,h=5)$. Indeed, spurious diffusion will occur first for the smallest microstructural length scales. Note that more elaborated techniques exist to circumvent spurious kinetic effects (see \cite{Tourret2021} and references therein). We choose $\kappa=10^{-5}$ since for $\kappa \geq 10^{-4}$, spurious diffusion occurs, whereas for $\kappa\leq10^{-6}$, the results are similar to the case where $\kappa=0$ (see Fig.~\ref{fig:effect_lap_coeff}). Specifically, it can be seen that for $\kappa \geq 10^{-4}$ the smallest pores close under diffusion of the grains phase, as opposed to the case where $\kappa \leq 10^{-5}$. The limit case $\kappa \gg \lambda$ of pure diffusion is also shown in Fig.~\ref{fig:effect_lap_coeff}, where, after only one timestep, the microstructure mixes into a homogeneous phase where $\varphi\approx0.6$. Thereupon, the mechanical response is that of a homogeneous elastic solid, which homogenized elastic modulus can be estimated around 6.

\subsection{Resemblance of the SMs with real microstructures}
\label{subsec:resemblence_with_real_microstructures}

\begin{figure}[h!]
	\centering
	\includegraphics[width=0.6\linewidth]{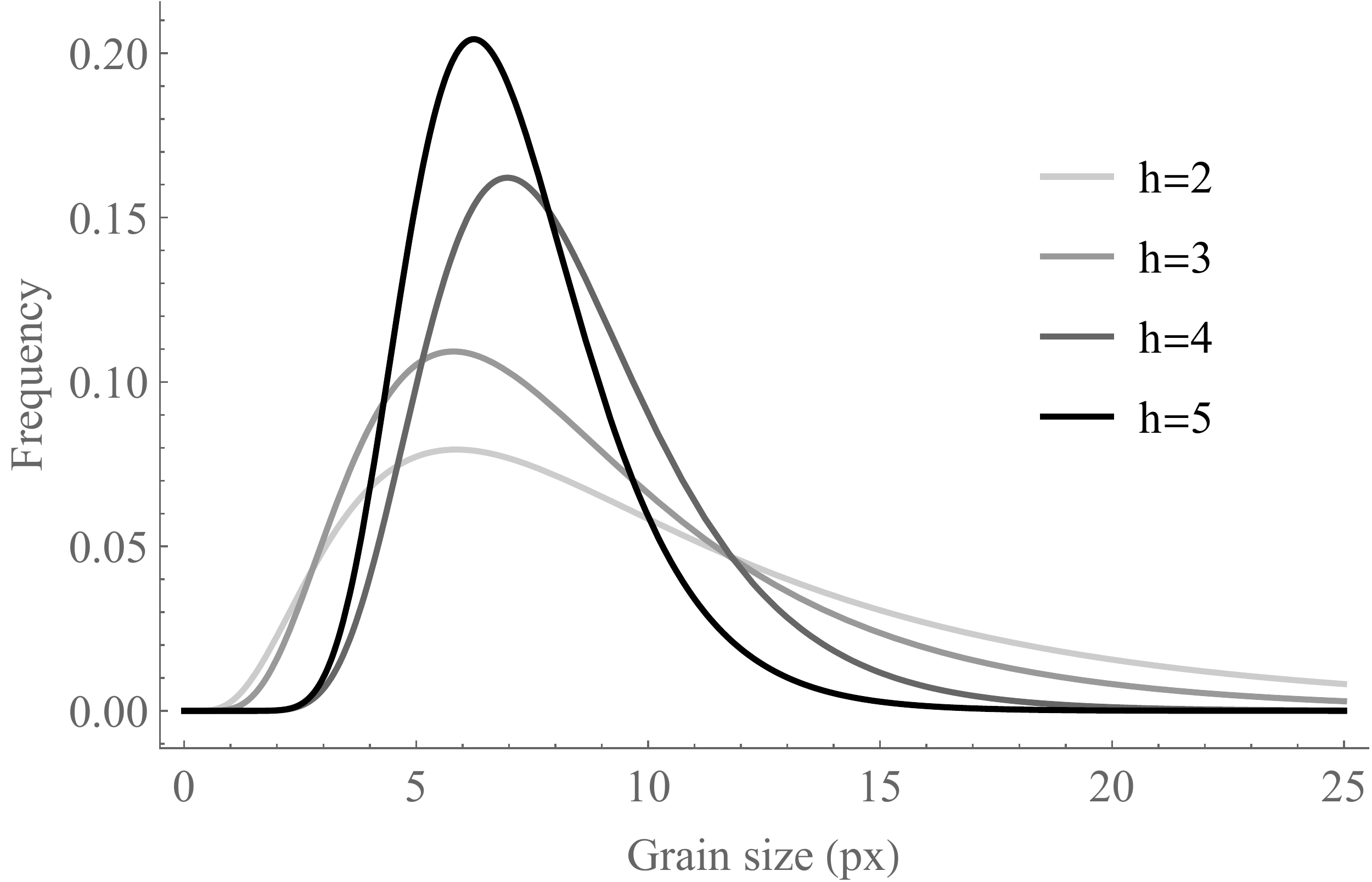}
	\caption{Evolution of the grain size distribution with the heterogeneity $h$ for a fixed porosity $n=0.2$ for a $200\times200$ pixels resolution, showing that higher values of $h$ lead to less skewed but narrower distributions. Indeed, the skewness values, calculated as Pearson's moment coefficient of skewness, are found here to be 2.75 for $h=2$, 1.96 for $h=3$, 1.07 for $h=4$, 0.95 for $h=5$.}
	\label{fig:GSDvsBlob}
\end{figure}

\begin{figure}[h!]
	\centering
	
		\begin{overpic}[width=0.21\linewidth]{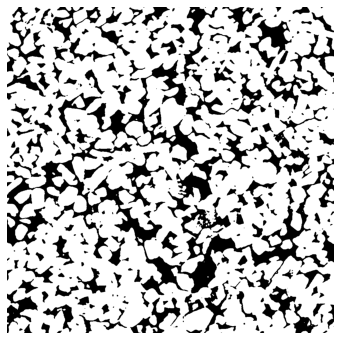}
		\put (-8,85) {\color{black} $a)$}
	\end{overpic}
	\hspace{0.02\linewidth}
	\begin{overpic}[width=0.21\linewidth]{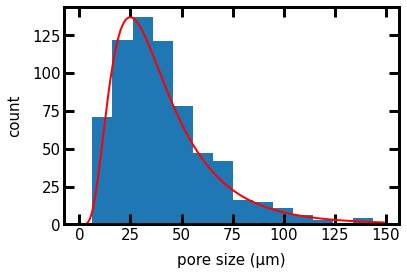}
		\put (1,65) {\color{black} $b)$}
	\end{overpic}
	\hspace{0.02\linewidth}
	\begin{overpic}[width=0.21\linewidth]{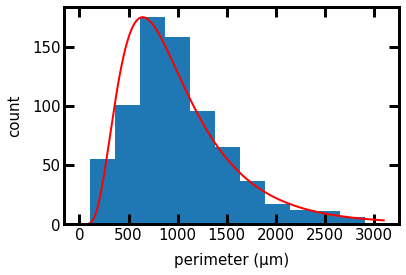}
		\put (1,65) {\color{black} $c)$}
	\end{overpic}
	\hspace{0.02\linewidth}
	\begin{overpic}[width=0.21\linewidth]{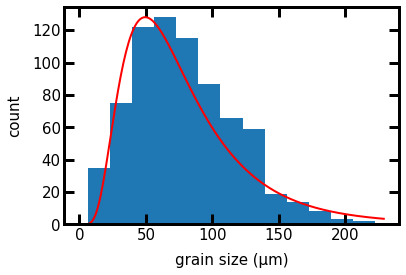}
		\put (1,65) {\color{black} $d)$}
	\end{overpic}
	
	\vspace{0.01\linewidth}
	
	\begin{overpic}[width=0.21\linewidth]{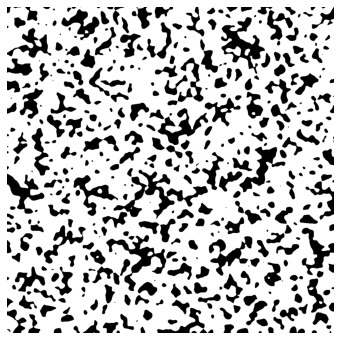}
	\put (-8,85) {\color{black} $e)$}
\end{overpic}
\hspace{0.02\linewidth}
\begin{overpic}[width=0.21\linewidth]{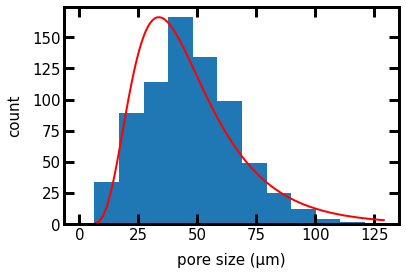}
	\put (1,65) {\color{black} $f)$}
\end{overpic}
\hspace{0.02\linewidth}
\begin{overpic}[width=0.21\linewidth]{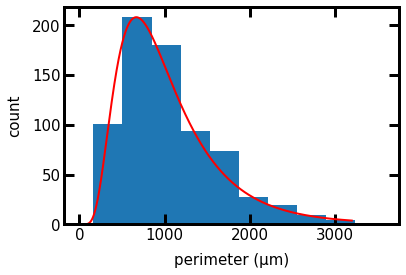}
	\put (1,65) {\color{black} $g)$}
\end{overpic}
\hspace{0.02\linewidth}
\begin{overpic}[width=0.21\linewidth]{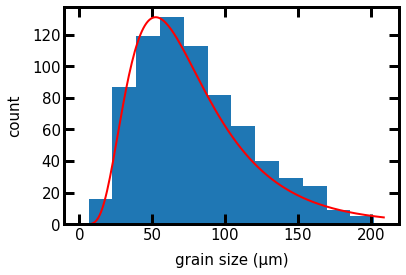}
	\put (1,65) {\color{black} $h)$}
\end{overpic}

	\caption{a) CT scan of Mt Simon sandstone segmented into 733 grains, of porosity $M_0=27.4\%$ (mean pore size $41.51 \mu m$), total perimeter $M_1=190.11mm$ (mean perimeter $1.02mm$) , mean grain size $M_3=79.80 \mu m$ and Euler number $M_4=-351$, along with the associated lognormally-fitted distributions in b), c), d) respectively. e) SM generated for $n=27.4\%$ and $h=3.9$ with resolution $1200\times1200$ pixels, matching the sandstone microstructure, segmented into 717 grains, of porosity $M_0=27.4\%$ (mean pore size $47.49 \mu m$), total perimeter $M_1=167.17mm$ (mean perimeter $1.06mm$) , mean grain size $M_3=79.26 \mu m$ and Euler number $M_4=-202$, along with the associated distributions in f), g), h) respectively.}
	\label{fig:resemblence_via_dist}
\end{figure}

Before drawing realistic results from the SMs, let us also check how representative they are of real microstructures. We find that the SMs resemble real microstructures through the distributions of the three first morphometers. We find that the grain size distributions of the SMs are skewed towards smaller grains (i.e.\ right-tailed), as expected from thresholding the initial symmetric Gaussian noise. The heterogeneity $h$ controls their skewness (i.e.\ deviation from a normal distribution) and narrowness (see Fig.~\ref{fig:GSDvsBlob}). The distributions of SMs with larger heterogeneity $h$ are less skewed but narrower. 
More precisely, the SMs exhibit lognormal distributions (see Fig.~\ref{fig:resemblence_via_dist}), similarly to a large variety of porous media. This type of distribution can be found in geomaterials (see \cite{Hwang2003}, \cite{Marks2015} and references therein), as well as engineered porous materials (see \cite{Liu2017} e.g.). For direct comparison, we consider a Mt Simon sandstone, presented in \cite{Kohanpur2020}, and digitalized on \textit{digitalrocksportal.org}, of average porosity $27.1\%$. Each sides of the CT scans have a resolution of 1200 pixels measuring each $2.8\mu m$. To imitate a given slice of porosity $27.4\%$ (Fig.~\ref{fig:resemblence_via_dist}a), we generate a SM (Fig.~\ref{fig:resemblence_via_dist}e) of same porosity and resolution, and of heterogeneity $h$ determined to match the mean grain size. The distributions of the three first morphometers for the real and synthetic microstructure are compared in Fig.~\ref{fig:resemblence_via_dist} and Tab.~\ref{tab:resemblance_comparison}, and checked to be both well fitted by lognormal distributions. 

\begin{table}[]
	\centering
	\begin{tabular}{|c|c|c|c|}
		\hline
		& \textbf{Sandstone} & \textbf{Synthetic} & \textbf{Error (\%)} \\ \hline
		Porosity $M_0$ (\%)            & 27.4               & 27.4               & 0                   \\ \hline
		Perimeter $M_1$ (mm)           & 190                & 167                & 12                  \\ \hline
		Mean grain size $M_2$ ($\mu$m) & 79.8               & 79.3               & 0.68                \\ \hline
		Mean pore size ($\mu$m)        & 41.5               & 47.5               & 14                  \\ \hline
		Mean perimeter (mm)            & 1.02               & 1.06               & 3.9                 \\ \hline
	\end{tabular}
\caption{Comparison of the morphometers of the Mt Simon sandstone with the synthetic analogue, along with the mean values of their lognormal distributions.}
	\label{tab:resemblance_comparison}
\end{table}


\subsection{Determinism of the SMs}
\label{subsub:determinism}

Since the generation of the SMs starts from a random noise, it is not a priori deterministic. Indeed, varying the random seeds for a given set of input parameters $n$ and $h$ yields different microstructures. To quantify the influence of this discrepancy on the mechanical response, we compare the numerical results for $n =0.3$, $h=3$ as an example, with 10 different seeds (see \ref{app:determinism_SM}). Namely, we calculate the set of flow stresses $\sigma_f$ (for $\epsilon_f=11.5\%$) for the different seeds. We find that the standard deviation of this set divided by its mean is around $4\%$, which we consider as an approximate measure of the potential error in the following predictive fitting.

\subsection{Morphometers evolution}
\label{subsec:numerical_output}
\begin{figure}[h!]
	\centering
    \includegraphics[width=1\linewidth]{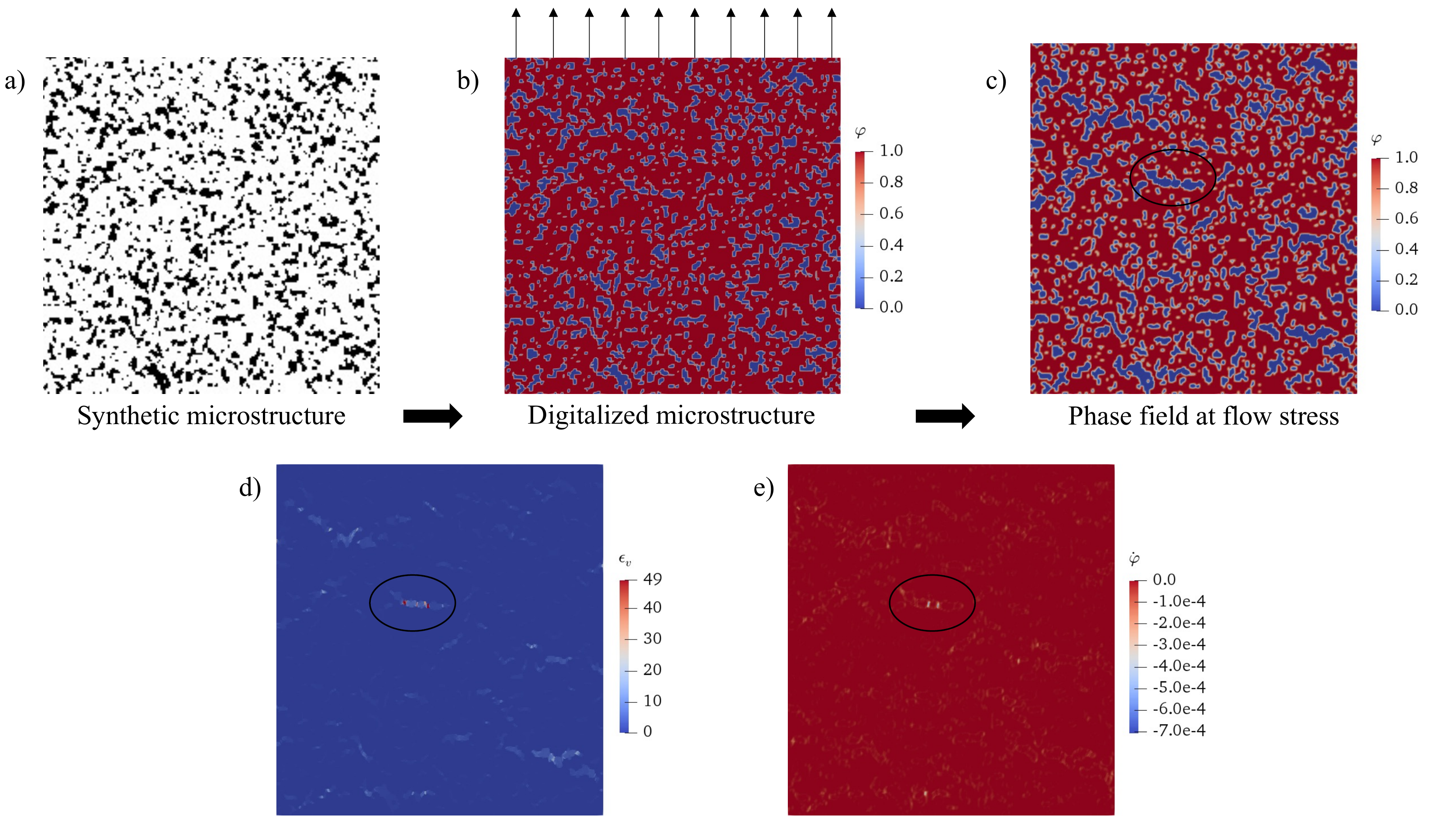}
	\caption{a) SM generated for $n=0.2$ and $h=5$. b) Digitization of the SM used as input for $\varphi$. c) Output for $\varphi$ after tensile loading post-yield at a certain flow stress for a vertical strain of $11.5\%$. d) Volumetric strain distribution for the same flow stress, concentrated in a central zone (black oval). e) Distribution of $\dot\varphi$ for the same flow stress, negative since the pore phase $\varphi=0$ is produced, with maximum absolute value also in the central zone, coinciding with the concentration of mechanical energy.}
	\label{fig:simu_output}
\end{figure}

\begin{figure}[h!]
	\centering
	\begin{overpic}[width=0.4\linewidth]{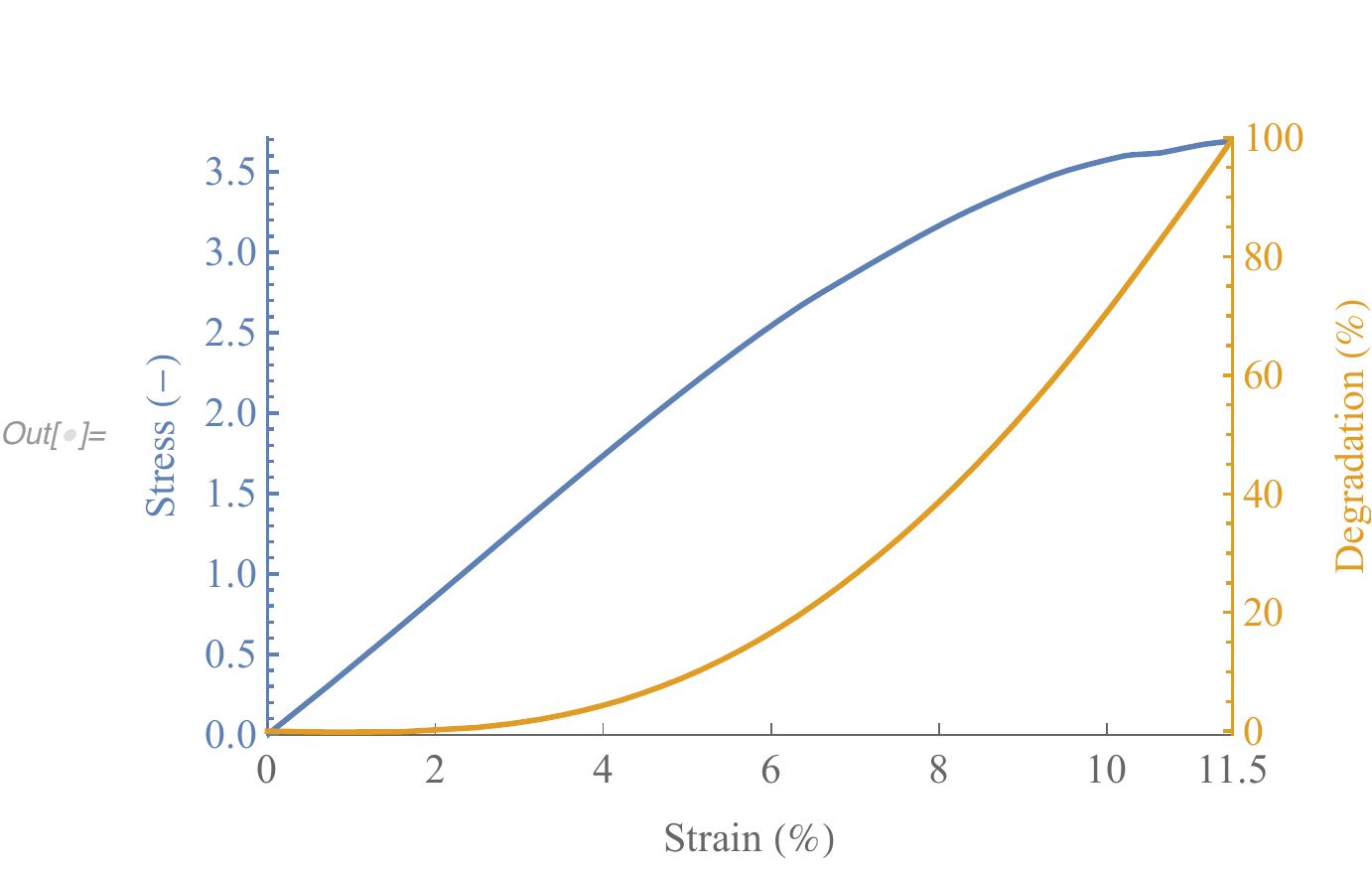}
		\put (-2,58) {\color{black} $a)$}
	\end{overpic}
	\hspace{0.02\linewidth}
	\begin{overpic}[width=0.4\linewidth]{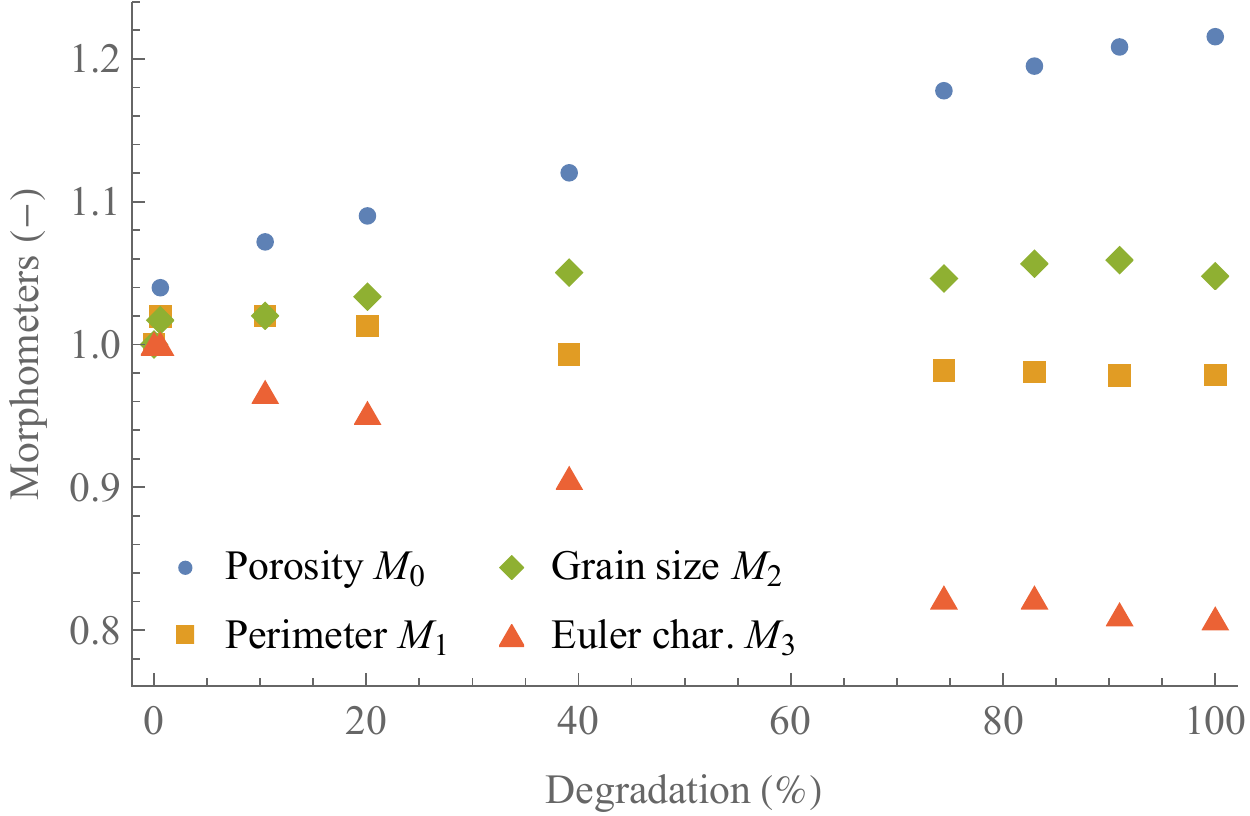}
		\put (-2,58) {\color{black} $b)$}
	\end{overpic}
	\caption{a) Stress-strain curve (in blue) corresponding to the evolution of the SM presented in Fig.~\ref{fig:simu_output}, along with the degradation-strain curve (in orange). b) Evolution of the morphometers with the degradation, whose values are normalized with their final values. Note that the values of the Euler characteristic $M_3$ is negative before normalization.}
	\label{fig:evolution_morphometers}
\end{figure}

Numerical results are shown in Fig.~\ref{fig:simu_output} for the SM $n=0.2$ and $h=5$, from the initial SM input, which digitization provides the initial conditions for the phase field $\varphi$, to the (post-yield) flow stress state. The associated distributions of the volumetric strain and of the phase-field rate illustrate the causality between mechanical loading and damage. Upon tensile loading, the pores open up, leading to damage events responsible for failure, namely the coalescence of neighboring pores, well known possible microstructural manifestation of strain hardening (\cite{Pardoen2000}). This can represent debonding or microcracks. In this particular example, the main damage event (maximum of $\lvert\dot\varphi\rvert$) responsible for the flow stress state occurs in the center (see black oval in Fig.~\ref{fig:simu_output}). 
As a macroscopic descriptor of the microscopic damage, we define the relative degradation $D_r$, similarly to breakage mechanics (\cite{Einav2007}), so that $D_r=0$ corresponds to an undamaged material and $D_r=100\%$ to a final state of damage. The latter is arbitrarily defined as the flow state (see Section \ref{subsec:measure_stength}). Thus, $D_r$ is a relative measure of degradation, and not a state variable, as it depends on the initial and final state considered, and is calculated as the following upscaling function:
\begin{equation}
	\label{eq:degradation}
	D_r=\frac{\int_{\bar\Omega} \varphi(\epsilon)\text{d}\epsilon-\int_{\bar\Omega} \varphi(\epsilon_0)\text{d}\epsilon}{\int_{\bar\Omega} \varphi(\epsilon_f)\text{d}\epsilon-\int_{\bar\Omega} \varphi(\epsilon_0)\text{d}\epsilon},
\end{equation}
where ${\bar\Omega}$ denotes the total spatial domain, $\epsilon_0=0$, $\epsilon_f$ is the strain when $\sigma_f$ is reached, so that $\int_{\bar\Omega} \varphi(\epsilon)\text{d}\epsilon$ computes the domain's portion occupied by the grains, since $\varphi=1$ characterizes the grains and $\varphi=0$ the pores. The evolution of the relative degradation of the SM of Fig.~\ref{fig:simu_output} with respect to the strain is shown in Fig.~\ref{fig:evolution_morphometers}, along with the corresponding stress-strain curve and the evolution of the 4 morphometers; although only 3 morphometers are required in 2D to fully constrain the strength law, we find it informative to include all of them here. As expected, the tensile loading opens up the pores, which increases the porosity $n=M_0$. Moreover, as the neighboring pores merge with each other towards failure (see Fig.~\ref{fig:simu_output}), the number of pores decreases (by around 9\%) and therefore, the Euler characteristic $\chi=M_3$ decreases in absolute value. As a consequence, the total perimeter $M_1$ decreases and the mean grain size $M_2$ increases, although both by less than 5\%, as opposed to $M_0$ and $M_3$ varying by around 20\%. The observed morphometers evolution is consistent with the evolution of strength found later in Section \ref{subsec:scaling_flow_stress} with respect to the initial morphometers values. Indeed, the material weakens upon loading, and therefore can be seen as a strong material initially and as a weaker one at the end of the loading. Consistently, we find that the strength decreases with the porosity $M_0$ and the mean grain size $M_2$, but increases with the total perimeter $M_1$ and the absolute value of the Euler characteristic $M_3$.

\section{Results}

\subsection{Scaling of flow stress with morphometers in SMs}
\label{subsec:scaling_flow_stress}

We now gather the results of all the simulations on the 42 SMs (see \ref{app:summary_table}), to find the best-fit function $f(M_i)$ relating the flow stress to the morphometers, via the least squares method. In the present 2D case, we only need 3 morphometers, chosen as $M_0$, $M_1$ and $M_2$ for instance. Following our initial assumption of a multiplicative dependence of the strength on the morphometers, as a generalization of Knudsen's theory (\cite{Knudsen1959}), we find that the best fit is exponential, with an adjusted $r^2$ coefficient of 0.9995 and an average error of 2.49\%, as compared with other potential candidates (see Tab.~\ref{tab:comparison_fitting}). In the present 2D case, the strength law Eq.~(\ref{eq:strength_law}) thus reads
\begin{equation}
	\label{eq:strength_law_2D}
	\sigma_f = \sigma_f^* e^{- \frac{M_0-M_0^*}{M_0^*}+\frac{M_1-M_1^*}{M_1^*}-\frac{M_2-M_2^*}{M_2^*}}.
\end{equation}
Notably, the fitting coefficient for the porosity $1/M_0^*=6.6$ is found to be in the range of values found experimentally by \cite{Knudsen1959}. The prediction is found to be as accurate when replacing $M_2$ by $M_3$. We check that a Hall-Petch type of fitting with respect to the mean grain size $M_2$ is also satisfying, while keeping the exponential dependence for $M_0$ and $M_1$. We find that the flow stress $\sigma_f$ increases with the total perimeter $M_1$ and with the absolute value of the Euler characteristic $M_3$. We also check that it decreases with the porosity $M_0$ and with the grain size $M_2$, as already well known. For a given random seed, the prediction is thus found to be erroneous by less than 3\% on average. Then, since choosing a seed amounts to an error of approximatively 4.33\% (see Section \ref{subsub:determinism}), we infer that the overall predictive error is $\sqrt{2.49^2+4.33^2}\approx5\%$. To use the latter formula, we have assumed that the  error propagation is multiplicative, since the choice of the seed and of the morphometry is concomitant. This error estimation relies, however, on assuming that the stochastic error of 4.33\% is valid for all the SMs, which is not here verified.
\begin{table}[]
	\centering
	\begin{tabular}{|c|c|c|c|c|c|}
		\hline
		Best fit   & \begin{tabular}[c]{@{}c@{}}Exponential\\ ($M_0$, $M_1$, $M_2$)\end{tabular}  & \begin{tabular}[c]{@{}c@{}}Exponential\\ ($M_0$, $M_1$, $M_3$)\end{tabular} & Power  & Hall-Petch & Linear \\ \hline
		Adjusted $r^2$     & 0.9995                            & 0.9995                            & 0.9974 & 0.9990     & 0.9780 \\ \hline
		Average error (\%) & 2.49                              & 2.47                              & 5.96   & 3.70       & 15.03  \\ \hline
		Fitting function &   $\alpha_1 e^{\alpha_2 M_0 +\alpha_3 M_1 +\alpha_4 M_2}$     & $\alpha_1 e^{\alpha_2 M_0 +\alpha_3 M_1 +\alpha_4 M_3}$      & $\alpha_1 M_0^{\alpha_2} M_1^{\alpha_3} M_2^{\alpha_4}$   & $\alpha_1 e^{\alpha_2 M_0 +\alpha_3 M_1}M_2^{-0.5}$       & $\alpha_1 M_0 +\alpha_2 M_1+\alpha_3 M_2$  \\ \hline
	\end{tabular}
\caption{Different fitting models with associated adjusted $r^2$ coefficient and average error, showing that the best fitting function is exponential.}	\label{tab:comparison_fitting}
\end{table}
The dependence of the strength on the porosity and grain size was already established to be exponential, as recovered here, or similarly, a power law (see \cite{Knudsen1959} and references therein). However, there are few results, if any, relating the strength to the two other morphometers, the surface area $M_2$ and the Euler number $M_4$, which could have corroborated our results.

\subsection{Dependence of the strength on the confinement}

\begin{figure}[h!]
	\centering
	\includegraphics[width=0.5\linewidth]{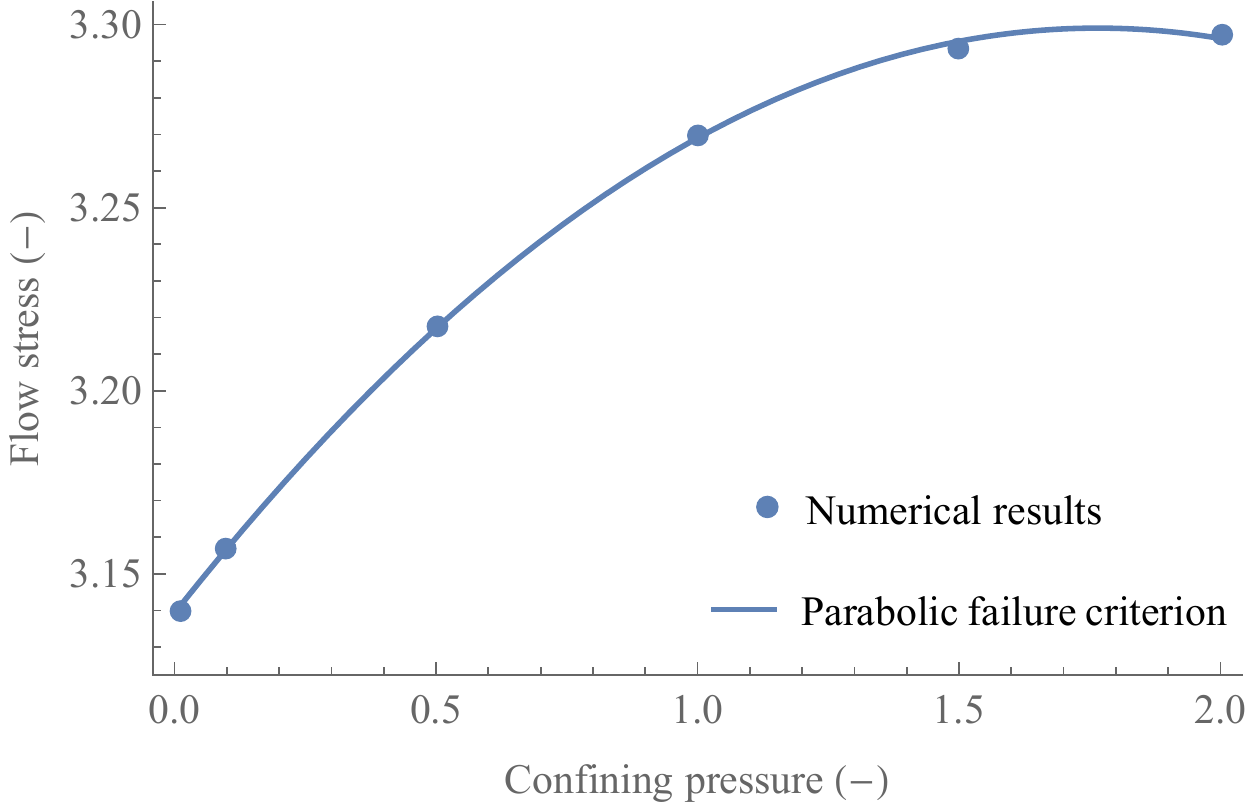}
	\caption{The points represent numerical results for the flow stress versus the confining pressure in axisymmetric biaxial compression. The curve corresponds to the fit $-0.0513p^2+0.181p+3.14$, for which the adjusted $r^2$ is exactly 1.}
	\label{fig:confinement}
\end{figure}

While the focus of this work concerns the morphometry, the strength can depend on other factors such as the confining pressure $P$, which we shall briefly discuss here. Such dependence can be included in the reference stress $\sigma_f^*$. We determine it by performing simulations of axisymmetric biaxial compressions for different values of $P$, for a given microstructure with values of porosity and heterogeneity in the middle range of our training batch ($n=0.3$, $h=3$), in plane strain conditions. To constrain the range of confining pressures, we estimate the preconsolidation pressure $P_c$ by performing an isotropic compression (see \ref{app:preconsolidation}). By reading the yield stress value in Fig.~\ref{fig:preconsolidation}, we find an approximate value of $P_c=2$. We then plot the obtained flow stresses with respect to the confining pressures (see Fig.~\ref{fig:confinement}), where the best fit function is a parabola. This recovers the parabolic failure criterion, recently suggested and experimentally verified (see \cite{Yuan2020}, \cite{Wang2019}, \cite{Singh2005}). This also captures the transition from a linear behavior at low confining pressures to a pressure-insensitive behavior for high confinement (\cite{Yuan2020}).

\subsection{Strength prediction from real microstructures}

To assess the validity of the morphometric strength law Eq.~(\ref{eq:strength_law_2D}), we predict the strength of real materials from CT scans, and compare it with phase-field simulations as previously. For that, we employ the following procedure. 1) We choose a set of $m$ CT scans $C^1,\dots,C^m$ of a given material and pick one of them as the reference $C^*$. 2) We binarize the $C^i$ and calculate the morphometers for each $C^i$ as done previously for the SMs. In particular, this determines the reference values for the morphometers $M_i^*$. 3) We run one simulation for the reference CT scan $C^*$ to determine the remaining reference value $\sigma_f^*$. 4) We deduce the $\sigma_f^i$ for the remaining CT scans using the strength law. 5) We run the simulations for the remaining CT scans for comparison. 

\subsubsection{Geomaterials}

\begin{figure}[h!]
	\centering
	\begin{overpic}[width=1\linewidth]{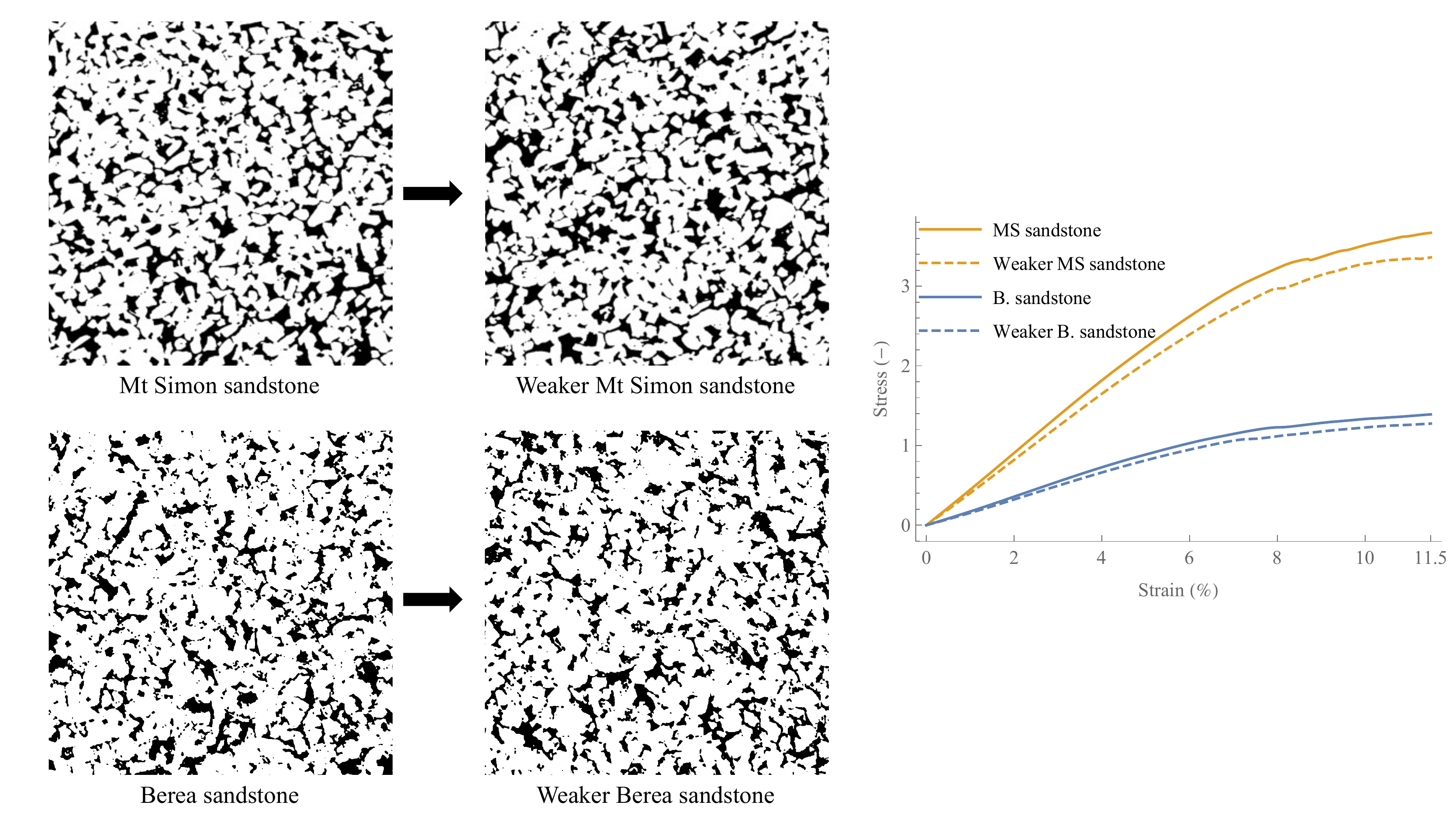}
		\put (0,52) {\color{black} $a)$}
		\put (31,52) {\color{black} $b)$}
		\put (0,24) {\color{black} $c)$}
		\put (31,24) {\color{black} $d)$}
		\put (60,40) {\color{black} $e)$}
	\end{overpic}
	\caption{a) CT scans of a Mt Simon sandstone (layer 108) and b) another layer of the same sandstone (890) with 8\% increase of porosity with similar other morphometers, hence weaker. c) a) CT scans of a Berea sandstone (layer 000) and b) another layer of the same sandstone (940) with 7\% increase of porosity with similar other morphometers, hence weaker. e) Mechanical response of the 4 sandstones microstructures.}
	\label{fig:sandstones}
\end{figure}
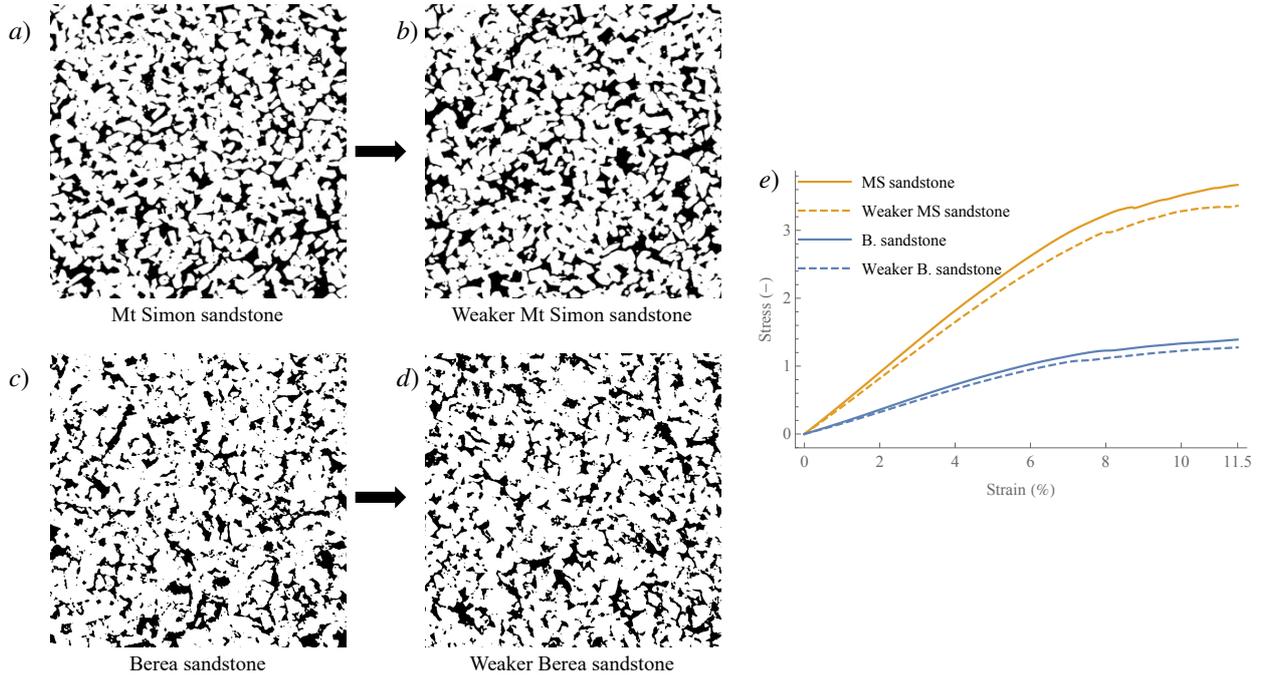

Starting with geomaterials, we use CT scans\footnote{digitalrocksportal.org} of Mt Simon and Berea sandstones, two benchmark rocks. The Mt Simon sandstone was introduced in Section \ref{subsec:resemblence_with_real_microstructures}. The Berea sandstone, of average porosity $17.9\%$, is presented in \cite{Lucas-Oliveira2020}, with a digitization resolution of $2.25 \mu m$ per pixel. The resolution of the CT scans of both sandstones is reduced to $300\times300$ pixels to reduce the mesh size and have reasonable computation times, in particular yielding erroneously high porosity values in Tab.~\ref{tab:prediction_sandstones_bones}. This does not hinder the present qualitative study since the different parts of the Berea sandstone remain comparable with each others. Using the same model Eq.~(\ref{eq:system_equations_dimensionless}) and parameters values, we follow the procedure described above. The results are gathered in Fig.~\ref{fig:master_fit} and detailed in Fig.~\ref{fig:sandstones} and Tab.~\ref{tab:prediction_sandstones_bones} for the example of a reference CT scan and a predicted flow stress for a weaker part with higher porosity. The predictions show a good agreement with an average error of 2.04\% for the Mt Simon sandstone and 3.88\% for the Berea sandstone. We must note that the choice of the reference CT scan $C^*$ was made judiciously to minimize this error. For instance, for the Mt Simon sandstone, while most choices of $C^*$ yielded an average error around 2 or 3\%, one in particular yielded an average error of 25\%. This seems to pertain to the usual problem in modeling geomaterials, that is, finding a representative elementary domain. Indeed, it is sensible that in order to deduce the properties of a material's domain, one must use a representative domain as a reference.

\subsubsection{Bones}

To showcase the general predictive power of the morphometric strength law Eq.~(\ref{eq:strength_law_2D}), we perform the same predictive procedure for bones, which are porous materials exhibiting a wide range of porosity and microstructural architecture different from that of geomaterials. In particular, the example of bones enables us to address anisotropic microstructures. Typical values of morphometers for geomaterials and bones are shown in Tab.~\ref{tab:prediction_sandstones_bones} for comparison. As far as the microstructure is concerned, there are two main kinds of bones, the cortical bone and the trabecular bone (see \cite{Cowin2002} for a general review of bone mechanics). The former occupies the perimeter of the bone and has a low porosity ranging typically from 5 to 25\%. The latter occupies the inside of the bone and has a scaffold microstructure, with a much higher porosity ranging from 50 to 90\%. The main morphometer varying from a bone to another is the porosity. Indeed, the creation of porosity, because of age or disease, is the main sign of bone degradation. Thus, we will use our strength law to predict the strength of bones with varying porosities, given a reference bone, typically a healthy bone with a normal porosity.

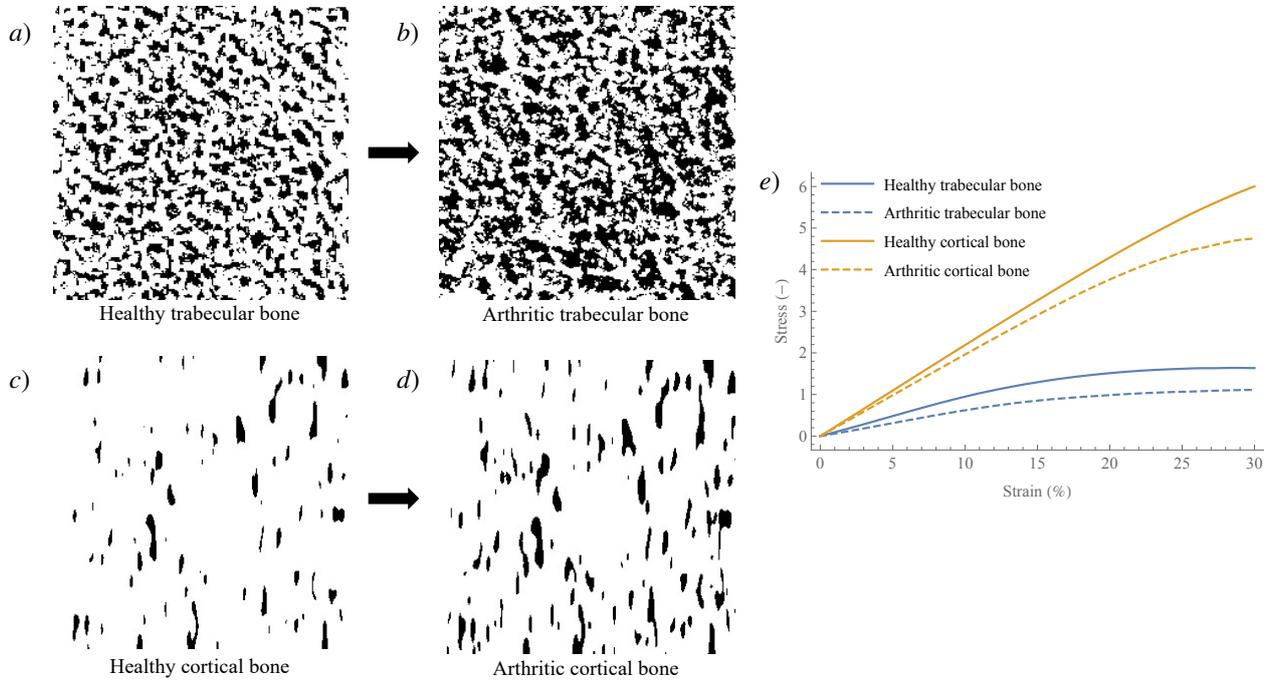
\begin{figure}[h!]
	\centering
	\begin{overpic}[width=1\linewidth]{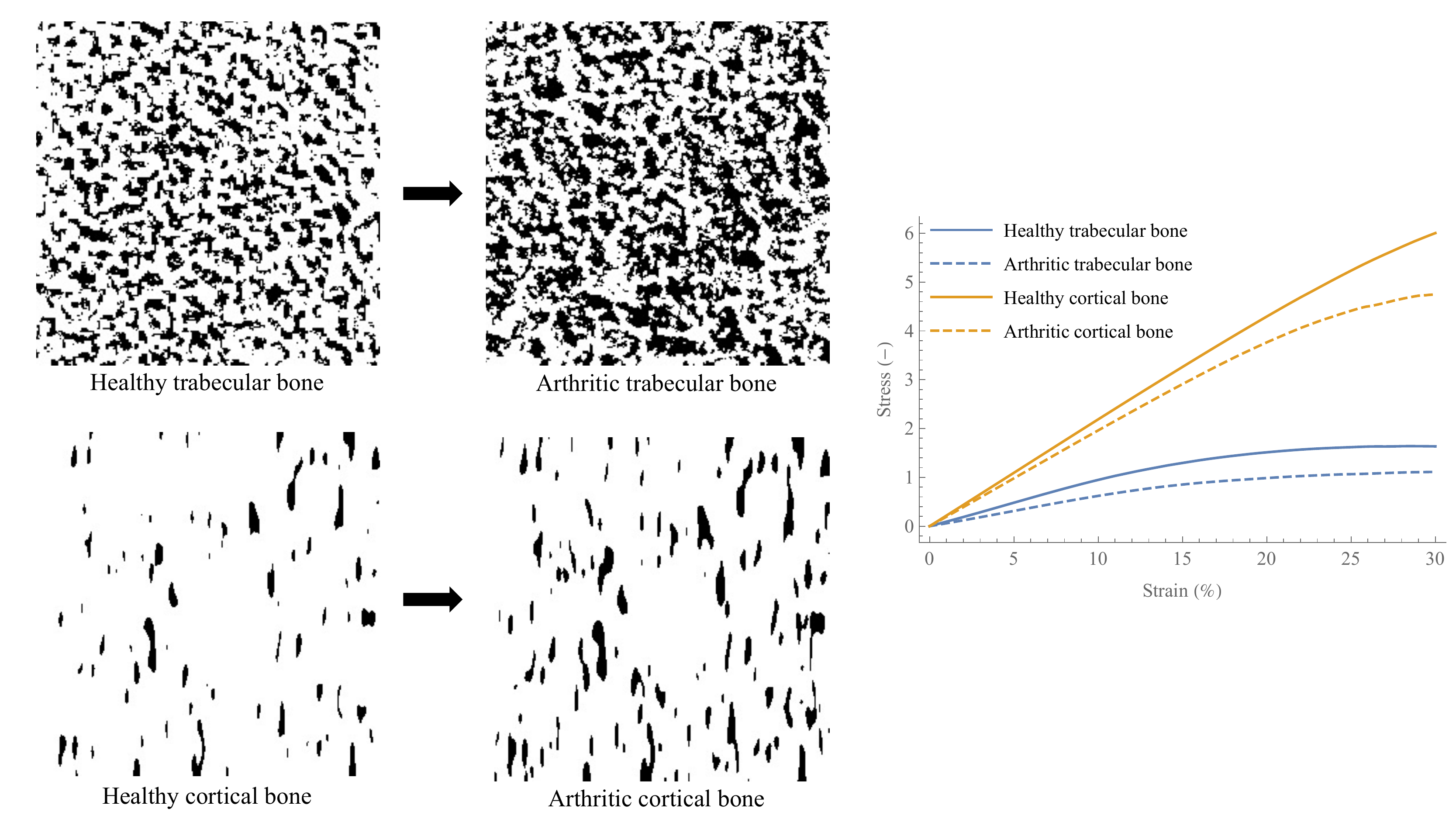}
		\put (-1,52) {\color{black} $a)$}
		\put (30,52) {\color{black} $b)$}
		\put (-1,24) {\color{black} $c)$}
		\put (30,24) {\color{black} $d)$}
		\put (59,40) {\color{black} $e)$}
	\end{overpic}
	\caption{a) CT scans of a healthy and b) arthritic trabecular bone's microstructure (from \cite{Neumann2018}). c) Synthetic microstructures of a healthy and d) arthritic cortical bone. e) Mechanical response of the 4 bones.}
	\label{fig:bones}
\end{figure}

For the trabecular bone, we used the CT scan in Fig.~1 of \cite{Neumann2018}, providing a transverse section. We choose the healthy microstructure as the reference, to predict the strength of the arthritic one (see Fig.~\ref{fig:bones}). As for the cortical bone, failing to find satisfying input images, we model it via SMs with anisotropy to resemble real cortical microstructures in their longitudinal section (see Fig.~9 in \cite{Cooper2004} and Fig.~1 in \cite{Granke2011}). For that, the heterogeneity $h$ is chosen as an array $(1,4)$, meaning that the pores are elongated in the vertical direction four times more than in the horizontal one (see Fig.~\ref{fig:bones}). We then generate different SMs with varying porosities (3, 5, 7, 10, 15 and 20\%) and fixed anisotropy. 

\begin{figure}[h!]
	\centering
	\includegraphics[width=0.7\linewidth]{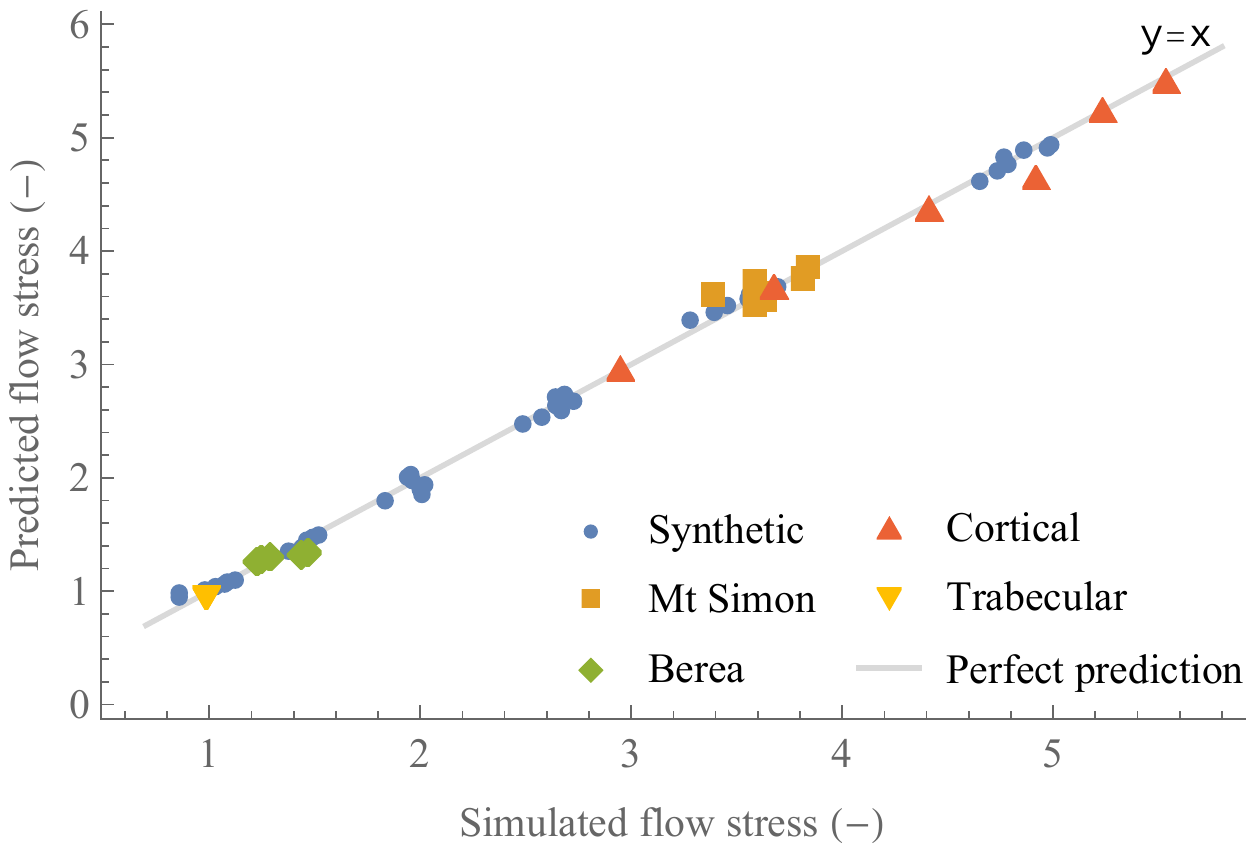}
	\caption{Predicted flow stress with the morphometric strength law Eq.~(\ref{eq:strength_law_2D}) versus the flow stress obtained from phase-field simulations, for SMs, geomaterials (Mt Simon and Berea sandstones) and biomaterials (trabecular bones and synthetic cortical bones).}
	\label{fig:master_fit}
\end{figure}

We use the same model as previously, but with different elastic moduli for the skeleton phase. The same problem of determining the moduli of the skeleton phase holds as for geomaterials. Indeed, elastic moduli are usually measured for the mixture of the skeleton and pores phases. As explained in Section (\ref{subsec: phase-field modeling}), we have assimilated the skeleton phase of geomaterials to a rock with very low porosity. Similarly, we assume the solid (mineral) phase of bones to be analogous to a bone of very low porosity, namely a cortical bone, for which the Lam\'e parameters can be found to be around $10GPa$ and $7GPa$ respectively (see \cite{Lai2015}). We choose the flow stress for the bones as $\sigma_f=25\%$. For the trabecular bone, since we use transverse sections, the loading is compressive. As for the cortical bone, modeled longitudinally, we choose a tensile loading. In agreement with \cite{Morgan2018}, we find that for a given post-yield vertical strain, the trabecular microstructure, in compression, has a softening response, whereas the cortical SM, in tension, is limited to a hardening response (see Fig.~\ref{fig:bones}). For the synthetic cortical bones, we show for instance the results for $n=5\%$ as a healthy bone and $n=10\%$ as the arthritic bone to be predicted, corresponding to real porosity values (see Fig.~6 in \cite{Pauly2015} for the lateral tibia). Finally, we show a relatively accurate prediction of the strength from our morphometric law (see Fig.~\ref{fig:master_fit}), with an error of 4.2\% for the trabecular bone (detailed in Tab.~\ref{tab:prediction_sandstones_bones}), and an average error of 1.3\% for the synthetic trabecular bones. We note that for the latter, the accurate prediction required using two different reference microstructures, one for $n<10\%$ and one for $n\geq10\%$, indicating that the material may have two distinct mechanical behaviors for those two ranges of porosities. We note also that we have used $M_3$ instead of $M_2$ for convenience, as determining the grain size distributions turns out less accurate for microstructures with notably low or high porosities and may not bear any physical meaning.
\begin{table}[h!]
	\centering
	\begin{tabular}{|c|c|c|c|c|c|c|c|}
		\hline
		&$M_0 (\%)$ & $M_1$ (mm) & $M_2$ ($\mu$m) & $M_3$ & $\sigma_f$ (numerics) & $\sigma_f$ (predicted) & Error (\%) \\ \hline
		Mt Simon sandstone   & 20.6 & 142   &118  & /     & 3.71      & *       & /      \\ \hline
		Weaker MS sandstone  & 22.2 & 145   &121  & /     & 3.39     & 3.59  & 6.09 \\ \hline
		Berea sandstone   & 37.7\~ & 253   &7.96  & /     & 1.33      & *       & /      \\ \hline
		Weaker B. sandstone  & 40.2\~ & 253   &8.35  & /     & 1.23     & 1.26  & 2.91 \\ \hline
		Healthy trabecular bone      & 29.5 & 451      & /         & -423  & 1.51      & *       & /      \\ \hline
		Arthritic trabecular bone   & 41.1   & 552     & /        &-265  & 0.986    & 0.944   & 4.2   \\ \hline
		Healthy cortical bone      & 5& 37.4      & /         & -67  & 5.23      & *       & /      \\ \hline
		Arthritic cortical bone   & 10   & 73.1     & /        &-91  & 4.41   & 4.36   & 1.08   \\ \hline
	\end{tabular}
	\caption{Details of the prediction of the strength of the Mt Simon and Berea sandstones and of the trabecular and cortical bones, for one reference and one prediction. *reference CT scans. \~\ Porosities larger than real values from lowering initial resolution.}
		\label{tab:prediction_sandstones_bones}
\end{table}
In all, the average errors between predicted and simulated flow stresses are all below 5\%, the expected maximum prediction error calculated in Section \ref{subsec:scaling_flow_stress}.

\section{Conclusion}
\label{}
We have suggested from qualitative 2D numerical simulations on synthetic and real microstructures the possibility to fully account for the morphometry of porous media in their macroscopic properties, such as the strength. To do so, we have used a damage phase-field modeling of the microstructure, capturing the exact microstructural geometry and encapsulating causes of damage such as debonding, dissolution and microcracks. The necessary and sufficient microstructural information is then upscaled in the form of morphometers defined from Minkowski functionals, as per Hadwiger's theorem. In the context of porous media, the four morphometers are chosen as the porosity, the total surface area of the skeleton, the mean grain size and the Euler characteristic; only three of them are required in 2D. We have inferred from a wide range of synthetic microstructures, with various porosities and heterogeneities, that the strength is best described by an exponential function of the morphometers. Those generated microstructures were checked to resemble real microstructures via the lognormal distribution of their morphometers. The exponential strength law was also checked to be accurate for two subclasses of porous media, geomaterials (sandstones) and biomaterials (bones), from digitalized CT scans. The strength prediction only requires the initial values of the morphometers for a given material's sample, along with the reference strength and morphometers values from another microstructure made of the same material.

The overarching goal is to predict the behavior of porous media with the minimal amount of data possible and in particular, minimizing the use of destructive tests in favor of remote sensing. This is of interest in geosciences where samples may be inaccessible and in biosciences where materials are best studied in vivo. The following steps in this endeavor will include 3D modeling and most importantly, experimental corroboration of the morphometric strength law. Then, the influence of environmental factors, like temperature and acidity, may be incorporated in a similar fashion. Building on aforementioned works, machine learning may also be used to ensure the realism of the synthetic microstructures. Ultimately, classical constitutive laws, such as viscoplasticity, may be re-explored to explicitly include the morphometric dependence. A major impediment to such program, albeit common to any modeling of heterogeneous microstructures, is that the reference sample must be representative of the material. This holistic description opens new avenues for modeling porous media in general, including geomaterials, biomaterials and engineered materials, at least as far as the microstructural geometry is concerned. It is indeed of paramount interest to draw bridges between the progress in different disciplines concerned with the same problematics.

\section{Acknowledgement}

The authors Alexandre Gu\'evel and Manolis Veveakis gratefully acknowledge the support of US DOE grant DE-NE0008746. 

 \appendix
 
 \newpage
  \section{Exponential relationship between strength and porosity for bones}
 \label{app:strength bones}
 \begin{figure}[h!]
 	\centering
 	\includegraphics[width=0.48\linewidth]{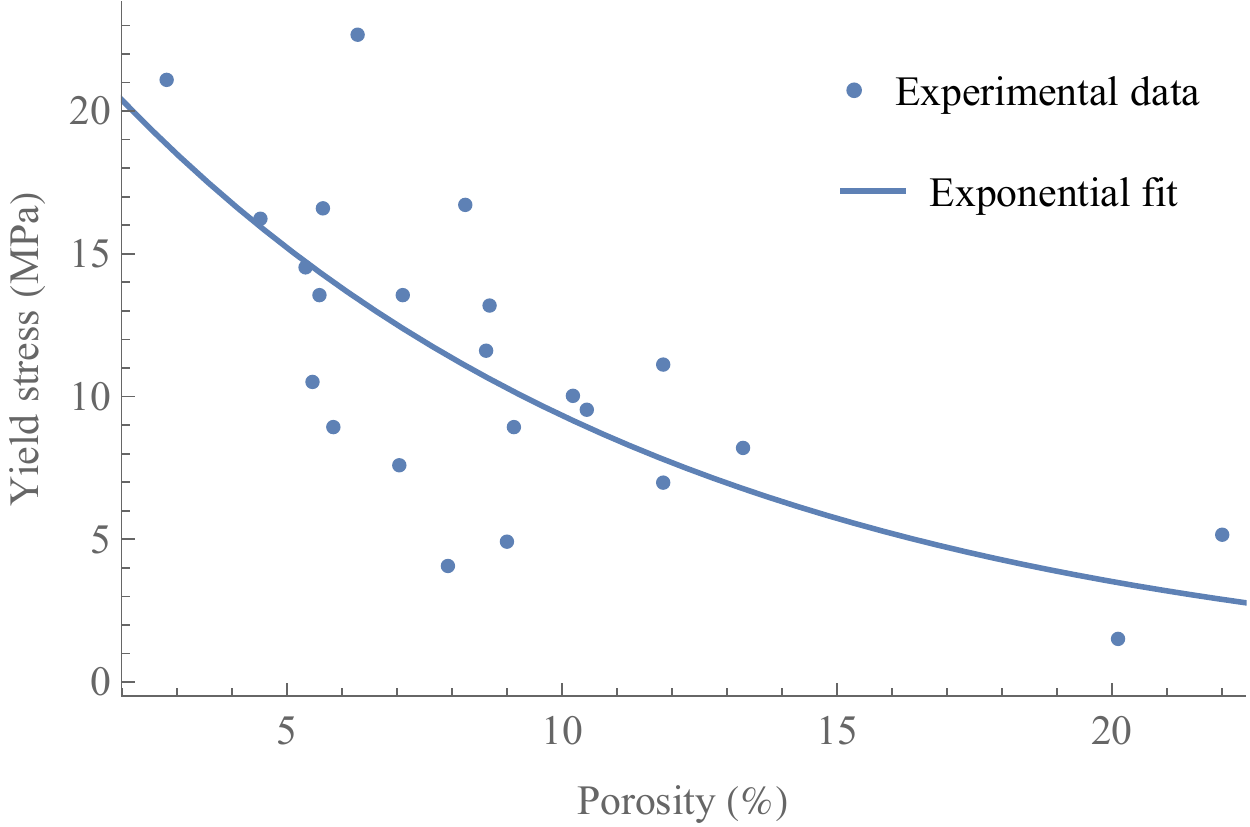}
 	\caption{Exponential fit of the yield stress evolution with porosity for cortical bones, adapted from the experimental results of \cite{Wachter2002}.}
 	\label{}
 \end{figure}

  \newpage
   \section{Summary table}
 \label{app:summary_table}

 \begin{figure}[h!]
 	\centering
 	\includegraphics[width=0.7\linewidth]{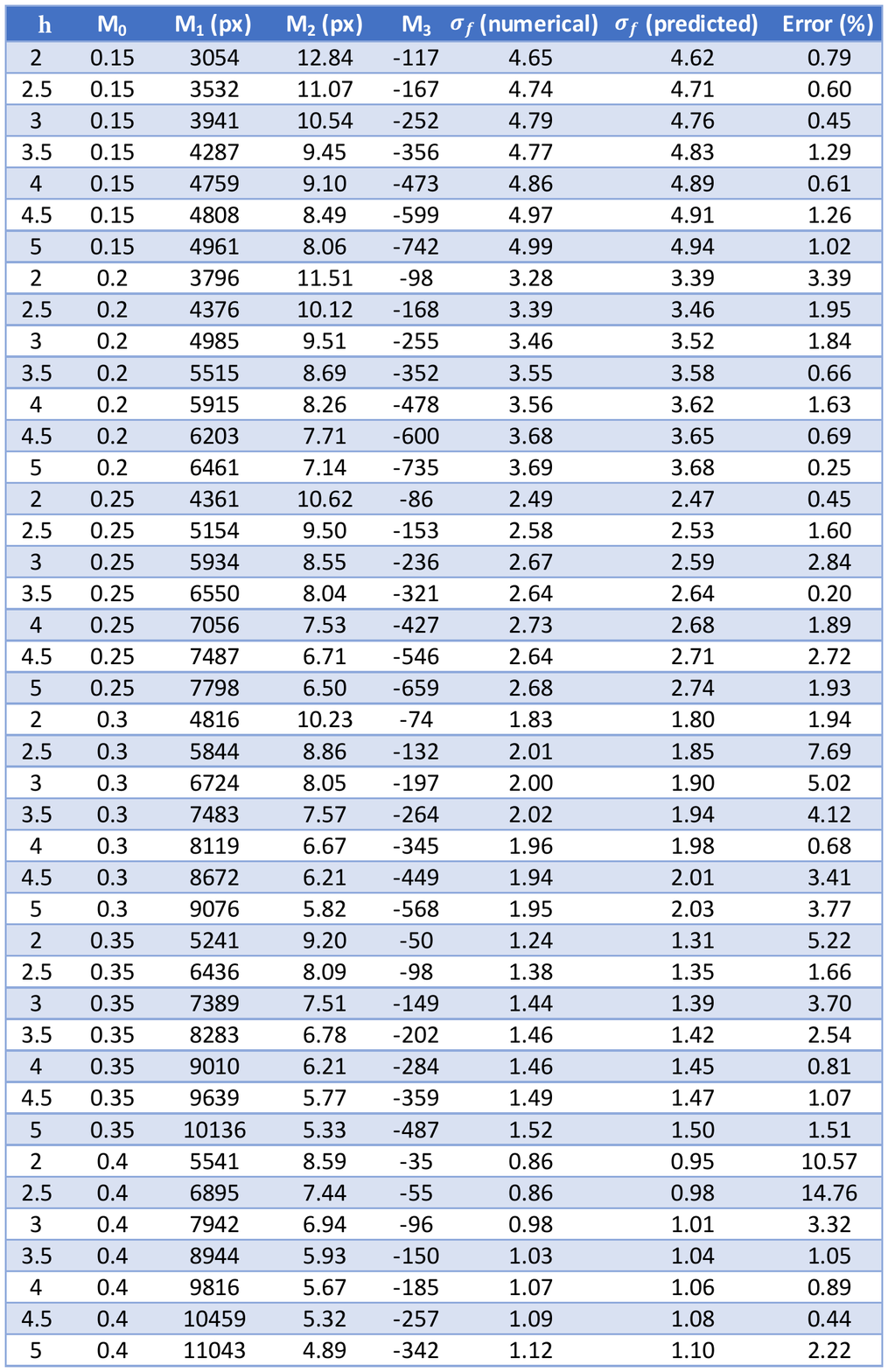}
 	\caption{Summary table gathering the morphometers of the SMs, with corresponding flow stresses obtained numerically with phase-field simulations, versus the flow stresses predicted with the morphometric strength law, along with the associated prediction error. The predicted flow stress is found to be best predicted with the exponential function $\sigma_f=12.3734 2^{-6.64471 e+0.938539p-0.951469g}$, obtained from \texttt{Mathematica}'s function \texttt{NonlinearModelFit}}
 	\label{fig:summary_table}
 \end{figure}

    \section{Calculations of the morphometers}
 \label{app:calculation_morphometers}
 
The morphometers are calculated via the Python libraries \texttt{PoreSpy} et \texttt{skimage}, as described in the following table. The segmentation of the SMs is performed via the SNOW algorithm introduced in \cite{Gostick2017}, using the function \texttt{filters.snow\_partitioning} in \texttt{PoreSpy}.

 \begin{table}[]
 	\centering
 	\begin{tabular}{|c|c|c|}
 		\hline
 		\textbf{Morphometer} & \textbf{Library}                  & \textbf{Function}           \\ \hline
 		Porosity $M_0$       & \texttt{PoreSpy} & \texttt{metrics.porosity}      \\ \hline
 		Perimeter $M_1$      & \texttt{skimage} & \texttt{measure.perimeter}     \\ \hline
 		Grain size $M_2$     & \texttt{PoreSpy} & \texttt{pore.diameter}*         \\ \hline
 		Euler number $M_3$   & \texttt{skimage} & \texttt{measure.euler\_number} \\ \hline
 	\end{tabular}
 \caption{* The function \texttt{pore.diameter} was initially defined (and named) to calculate the sizes of the pores obtained from segmentation. However, it can be applied to calculate the grain size when applied on the segmentation of the grains.}
 \end{table}

   \section{Determinism of the SMs}
 \label{app:determinism_SM}

 \begin{figure}[h!]
 	\centering
 	\includegraphics[width=0.7\linewidth]{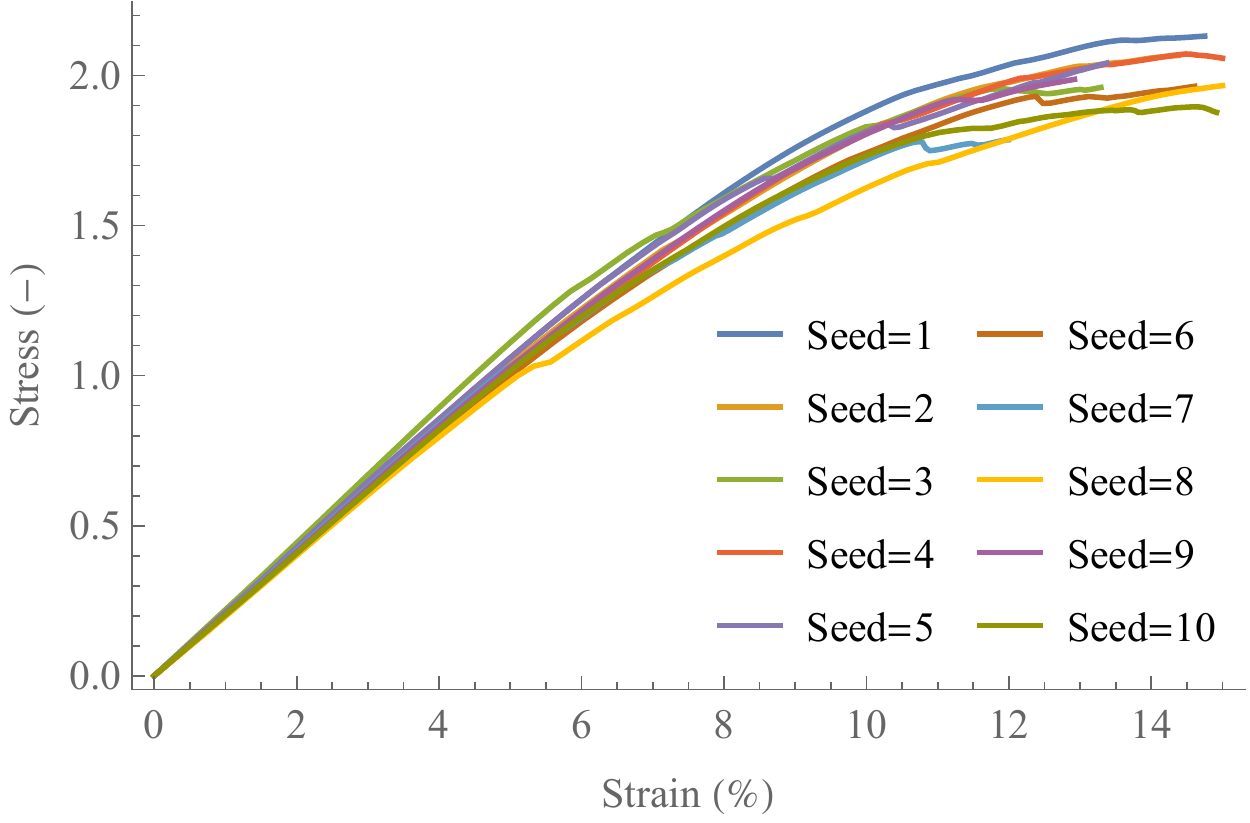}
 	\caption{Mechanical response for different random seeds for the SM $n=0.3$, $h=3$.}
 	\label{fig:determinism_SM}
 \end{figure}

  \newpage
  \section{Preconsolidation}
 \label{app:preconsolidation}

 \begin{figure}[h!]
 	\centering
 	\includegraphics[width=0.5\linewidth]{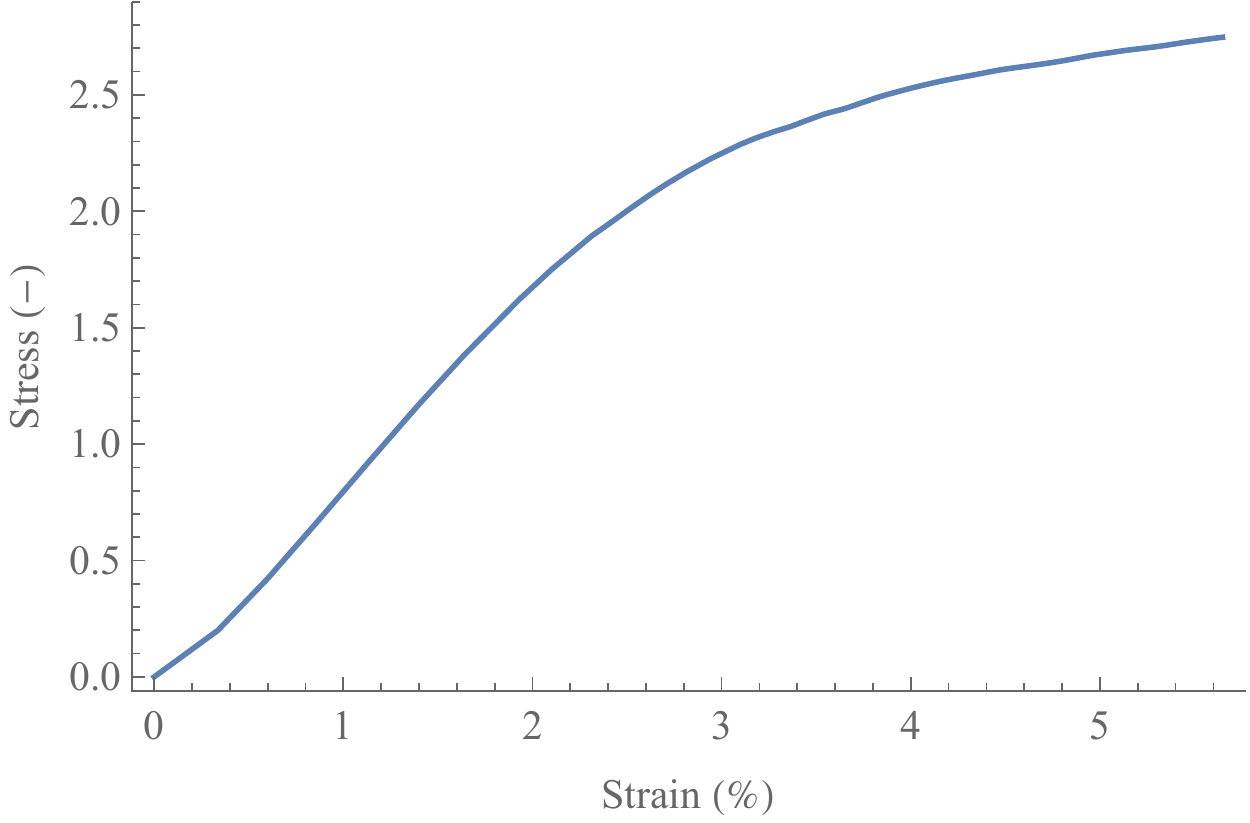}
 	\caption{Stress-strain curve of the isotropic compression simulation determining the preconsolidation stress, taken as the yield stress of this curve and found to be around 2.}
 	\label{fig:preconsolidation}
 \end{figure}

\newpage

\section*{References}
\bibliographystyle{elsarticle-harv} 
\bibliography{JMPS2021}

\end{document}